\documentclass[published]{JHEP3} % 10pt is ignored!

\JHEP{00(2005)000}

\JHEPspecialurl{http://jhep.sissa.it/JOURNAL/JHEP3.tar.gz}

\usepackage{epsfig,multicol,amsmath,amssymb,amsfonts}

\newcommand\fverb{\setbox\pippobox=\hbox\bgroup\verb}
\newcommand\fverbdo{\egroup\medskip\noindent%
            \fbox{\unhbox\pippobox}\ }
\newcommand\fverbit{\egroup\item[\fbox{\unhbox\pippobox}]}
\newbox\pippobox
\title{Infinite-dimensional representations of the rotation group and Dirac's monopole problem}

\author{Alexander I. Nesterov and F. Aceves de la Cruz\\
  E-mail: \email{nesterov@cencar.udg.mx}, \email{fermin@udgphys.intranets.com}\\

    Departamento de F{\'\i}sica, CUCEI, Universidad de
Guadalajara, Av. Revoluci\'on 1500, Guadalajara, CP 44420, Jalisco,
M\'exico\\
} \received{}        %%
\revised{\today}
\accepted{}        %% These are for published papers.

\preprint{\hepth{9912999}}  % OR: \preprint{Aaaa/Mm/Yy\\Aaa-aa/Nnnnnn}
\abstract{Within the context of infinite-dimensional representations of the
rotation group the Dirac monopole problem is studied in details. Irreducible
infinite-dimensional representations, being realized in the indefinite metric
Hilbert space, are  given by linear unbounded operators in infinite-dimensional
topological spaces, supplied with a weak topology and associated weak
convergence. We argue that an arbitrary magnetic charge is allowed, and the
Dirac quantization condition can be replaced by a generalized quantization rule
yielding a new quantum number, the so-called {\em topological spin}, which is
related to the weight of the Dirac string.

 }


\begin{document}

\section{Introduction}

In 1931 Dirac \cite{Dir} showed that a proper description of the quantum
mechanics of a charged particle of the charge $e$ in the field of the magnetic
monopole of the charge $q$ requires the quantization condition $2\mu =n, \, n
\in \mathbb Z$ (we set $eq =\mu$ and $\hbar = c =1$). But in spite of numerous
efforts, Dirac monopole was not found so far.

Recently renewed interest in the Dirac monopole has been grown in connection
with the `fictitious' monopoles that are similar to the `real' magnetic
monopoles, however, appearing in the context of the Berry phase \cite{Berry}.
These type of magnetic monopoles emerge in the anomalous Hall effect of
ferromagnetic materials, trapped $\Lambda$-type atoms, anisotropic spin
systems, noncommutative quantum mechanics,  etc., and {\em may carry an
arbitrary `magnetic' charge} \cite{Br,ZLS,Hal,FP,SR,MSN,BM,FNT}.

Widely accepted group theoretical, topological and geometrical arguments in
behalf of Dirac quantization rule are based on employing classical fibre bundle
theory or finite dimensional representations of the rotation group
\cite{Wu1,Wu2,Jac,Gr,Gr1,G1,G2,G3}. For instance, a realization of the Dirac
monopole as U(1) bundle over $S^2$ implies that there exists the division of
space into overlapping regions $\{U_i\}$ with nonsingular vector potential
being defined in $\{U_i\}$ and yielding the correct monopole magnetic field in
each region. On the triple overlap $U_i\cap U_j\cap U_k$ it holds
\begin{equation}
\exp(i(q_{ij}+q_{jk}+q_{ki})) =\exp(i4\pi\mu)
\end{equation}
where $q_{ij}$ are the transition functions, and the consistency condition,
which is equivalent to the associativity of the group multiplication, requires
$q_{ij}+q_{jk}+q_{ki} = 0 \mod2\pi \mathbb Z$. This yields the Dirac
selectional rule $2\mu \in \mathbb Z$ as a necessary condition to have a
consistent U(1)-bundle over $S^2$ \cite{Wu1,Wu2,Gr}. Thus to avoid the Dirac
restrictions on the magnetic charge one needs to consider a {\em
nonassociative} generalization of U(1) bundle over $S^2$.  Recently we have
developed a consistent pointlike monopole theory with an arbitrary magnetic
charge in the framework of nonassociative fibre bundle theory
\cite{N1,N2,NF,N1a}.

As is known, in the presence of the magnetic monopole the operator of the total
angular momentum ${\mathbf J}$, which includes contribution of the
electromagnetic field, obeys the standard commutation relations of the Lie
algebra of the rotation group
\begin{eqnarray*}
[J_i, J_j] = i\epsilon_{ijk}J_k \label{eq4a}.
\end{eqnarray*}
These commutation relations fail on the Dirac string restricting the domain of
definition of the operator $\mathbf J$ and limiting it to the functions that
vanish sufficiently rapidly on the string \cite{Zw_1,Str}. In fact, any
approach to the charge quantization uses some additional assumptions, and in
group theoretical treatment this is the {\em requirement that $J_i$'s generate
a finite-dimensional representation of the rotation group yielding } $2\mu \in
\mathbb Z$ \cite{Ch,Gol1,Gol2,Zw_1,Zw_2,H}. Thus, one should give up
finite-dimensional representations of the rotation group to allow an arbitrary
magnetic charge.

Here we study in details the Dirac monopole problem within the framework of
infinite-dimensional representations of the rotation group. The paper is
organized as follows. In Section II the indefinite metric Hilbert space is
introduced. In Section III the properties of infinite-dimensional
representations of the rotation group are discussed. In Section IV it is argued
that extending the representations of the rotation group to
infinite-dimensional representations allows an arbitrary magnetic charge. In
Section V the obtained results and open problems are discussed.

\section{Indefinite metric Hilbert space}

Starting from the early 1940s indefinite metric in the Hilbert space has been
discussed and used by many authors. Recently a growing interest to this topic
has been risen in the context of the so-called PT-symmetric quantum mechanics
related to some non-Hermitian Hamiltonians with a real spectrum and
pseudo-Hermitian operators \cite{BB,BBM,BBJ,M1,M2,M3,M4,M5,Sol,BSS,RM}.

In conventional quantum mechanics the norm of quantum states given by
\begin{equation}\label{n}
 \int \bar\psi \psi d x > 0
\end{equation}
where $\bar \psi$ is the conjugate complex of $\psi$, carries a probabilistic
interpretation, and appearance of an indefinite metric in Hilbert space is a
sever obstacle. In particular, this leads  to {\em negative probability of
states}, that means observables with only positive eigenvalues can get negative
expectation values \cite{Pauli}.

We treat here more general situation when the normalization given by
\begin{equation}\label{n1}
 \int \bar\psi \psi d \mu(x),
\end{equation}
$d \mu$ being a suitable measure, is not necessary positive. We assume
that the integral
\begin{equation}\label{n1a}
 \int \bar\psi \psi' d \mu(x),
\end{equation}
may be divergent and its value is given by a regularization (for the definition
of regularization of an integral see, {\em e.g.} \cite{GSH}). There exist
several possibilities of regularizing divergent integral, further  we consider
the regularization of the integral by analytical continuation in parameter
(Sect. III), and the regularized integral will be denoted by
${\smallsetminus{\hspace{-0.3cm}}\int}$.

Following the notations introduced by Pauli \cite{Pauli}, we consider an {\em
inner product} in the indefinite metric Hilbert space $\mathcal H^\eta$ defined
by the bilinear form of the type
\begin{equation}
(\psi,\psi')_\eta = (\psi,\eta\psi') ={\smallsetminus{\hspace{-0.36cm}}\int}
\bar\psi {\eta} \psi' d \mu(x), \label{h1}
\end{equation}
in which the operator $\eta$ is only restricted by the condition that it
has to be Hermitian and
\begin{equation}\label{n2}
 {\smallsetminus{\hspace{-0.36cm}}\int} \bar\psi {\eta} \psi d \mu(x) > 0.
\end{equation}
The difference between our construction of the indefinite metric
Hilbert space and one been suggested in \cite{Pauli} arises from the
restriction (\ref{n2}). While Pauli requires the positive defined
norm (\ref{n1}), we don't.

Let functions $\psi_m(x)$ form the basis such that
\begin{eqnarray*}
{\smallsetminus{\hspace{-0.36cm}}\int} \overline \psi_m(x)\psi_{m'}(x)d\mu(x)=
\eta_{mm'}
\end{eqnarray*}
where  $\eta_{m m'}= (-1)^{\sigma(m)}\delta_{mm'}$ is an indefinite diagonal
metric, and $(-1)^{\sigma(m)} = \pm 1$ depending whether $\sigma(m)$ is even or
odd. Defining the action of the operator $\eta $ on $\psi_m$ as
\begin{equation}\label{eta1}
    \eta\psi_m = \eta_{mm'}\psi_{m'}
\end{equation}
we find that the set $\{\psi_m\}$ forms the orthonormal basis with respect
to the inner product given by
\begin{equation}\label{eta2}
(\psi_m,\psi_{p})_\eta =  {\smallsetminus{\hspace{-0.36cm}}\int} \overline
\psi_m(x)\eta_{pm'}\psi_{m'}(x)d\mu(x) =\delta_{mp}.
\end{equation}

Since the set $\{\psi_m(x)\}$ forms a basis, an arbitrary function
$\psi(x)\in\mathcal H^\eta$ can be expanded in terms of the
$\psi_m(x)$:
\begin{eqnarray}
\psi(x) = \sum_m c^\eta_m \psi_m,
\end{eqnarray}
where
\begin{equation}\label{N2}
c^\eta_m = (\psi_m, \psi)_\eta =
\eta_{mm'}{\smallsetminus{\hspace{-0.36cm}}\int} \overline
\psi_{m'}(x)\psi(x)d\mu(x)
\end{equation}

Let
\begin{align}
    &\psi'(x) = \sum_m c'^\eta_m \psi_m,
\end{align}
then the inner product $(\psi,\psi')_\eta$ can be easily calculated
and the result is:
\begin{equation}
(\psi,\psi')_\eta = {\smallsetminus{\hspace{-0.36cm}}\int} \bar
\psi(x){\eta}\psi'(x)d\mu(x) = \sum_m \overline c^\eta_m c'^\eta_m. \label{h3}
\end{equation}
In particular, one has
\begin{equation*}
 (\psi,\psi)_\eta = {\smallsetminus{\hspace{-0.36cm}}\int} \bar \psi(x){\eta}\psi(x)d\mu(x) = \sum_m |c^\eta_m|^2 > 0.
\end{equation*}
Thus  we see that the inner product in the indefinite metric Hilbert space is
positive defined scalar product. This provides the standard probabilistic
interpretation of the quantum mechanics.

The inner product ({\ref{h3}}) may be written in another form. Let us
consider the sum
\begin{equation}\label{k_3}
K(x,x') = \sum_m \psi_m(x)\overline {\psi_m(x')}.
\end{equation}
This yields the following relations:
\begin{align}
 & {\smallsetminus{\hspace{-0.36cm}}\int} \psi_{m}(x')K(x,x') d\mu(x')= \eta_{m m'}\psi_{m'}(x),
  \label{k_2}\\
&{\eta}\psi'(x)=  {\smallsetminus{\hspace{-0.36cm}}\int}
K(x,x')\psi'(x')d\mu(x'), \label{k_2a}
\end{align}
and it is seen that the kernel $K(x,x')$ plays here a role similar to that of
$\delta$-function in the standard Hilbert space of quantum mechanics. Now one
can rewrite the inner product (\ref{h3}) as
\begin{equation}
(\psi,\psi')_\eta = {\smallsetminus{\hspace{-0.36cm}}\int}
{\smallsetminus{\hspace{-0.36cm}}\int}\bar
\psi(x)K(x,x')\psi'(x')d\mu(x)d\mu(x').
\end{equation}

The expectation value of an observable $A$ represented by the linear
operator acting in $\mathcal H^\eta$ is defined by
\begin{equation}\label{A1}
    \langle A\rangle_\eta ={\smallsetminus{\hspace{-0.36cm}}\int} \bar \psi(x){\eta}A\psi(x)d\mu(x),
\end{equation}
and a generalization of the Hermitian conjugate operator, being denoted as
$A^\dag_\eta$, is given by
\begin{equation}\label{A2}
A^\dag_\eta = \eta^{-1}A^\dag \eta
\end{equation}
where $A^\dag$ is the Hermitian conjugate operator.

Since the observables are real, we see that the related operators have to
be self-adjoint in the indefinite metric Hilbert space, that means
$A^\dag_\eta= A$. In particular, applying this to the Hamiltonian operator
$H$, we have $ H^\dag_\eta= H$, and assuming that the wave function
satisfies the Schr\"odinger's equation
\[
i\frac{\partial{\psi}}{\partial{t}} = H\psi,
\]
we obtain the conservation of the wave function normalization:
\begin{equation}\label{A3}
    \frac{d}{dt}(\psi,\psi)_\eta = i{\smallsetminus{\hspace{-0.36cm}}\int}\bar\psi \eta(H^\dag_\eta- H)\psi d\mu(x) =0.
\end{equation}

Let us perform a linear transformation  $$\psi = S\psi',$$ then in
order to conserve the normalization of the wave function
\[
(\psi',\psi')_\eta = (\psi,\psi)_\eta
\]
one has to demand
\begin{equation}\label{A4}
    \eta' = S^\dag\eta S.
\end{equation}
In a similar manner we find that the observables are invariant,
\begin{equation}\label{A5}
\langle A'\rangle_\eta =\langle A\rangle_\eta,
\end{equation}
if the operators transform as follows:
\begin{equation}\label{A5a}
    A'= S^{-1}A S, \quad A^{\dag'}_\eta= S^{-1}A^\dag_\eta S.
\end{equation}

Assuming that, according to (\ref{A5a}), the matrix $A$ can be transformed with
a suitable $S$ to a normal form  such that
\begin{equation}\label{A6}
A\psi_n = a_n\psi_n
\end{equation}
we  find
\begin{equation}\label{A6a}
(\psi,A\psi)_\eta = \sum_n a_n|c^\eta_n|^2.
\end{equation}
This leads to the conclusion that {\em operator with only positive eigenvalues
can not have negative expectation values}. In other words in our approach, in
contrast to the others recently have been developed in the growing number of
papers on the subject of $PT$-symmetric quantum mechanics, the negative
probabilities do not appear and the standard probabilistic interpretation of
the quantum mechanics is preserved.

\section{Infinite-dimensional representations of the rotation group}

The three dimensional rotation group is locally isomorphic  to the group
$SU(2)$, and as well known $SO(3)= SU(2)/\mathbb{Z}_2$. In what follows
the difference between $SO(3)$ and $SU(2)$ is not essential and actually
we will consider  $G=SU(2)$. The Lie algebra corresponding to the Lie
group $SU(2)$ has three generators and we adopt the basis $J_{\pm} = J_{1}
\pm iJ_{2}, \; J_3$.  The commutation relations are:
\begin{align}
&[J_{+}, J_{-}] = 2J_3, \quad [J_3, J_{\pm}] = \pm J_{\pm}\label{J_01} \\
 &[J^2 , J_{\pm}]=0, \quad [J^2,J_3] =0
 \label{J_1}
\end{align}
where
\begin{eqnarray}
J^2=J_3^2 +\frac{1}{2}(J_{-}J_{+}+J_{+}J_{-}) \label{Cas}
\end{eqnarray}
is the Casimir operator.

Let $\psi^\lambda_\nu$ be an eigenvector of the operators $J_3$ and $J^2$
such that
\begin{equation}\label{J1a}
J_3\psi^\lambda_\nu=(\nu +n)\psi^\lambda_\nu,\quad J^2\psi^\lambda_\nu
=\lambda(\lambda +1)\psi^\lambda_\nu,
\end{equation}
where $n=0,\pm 1,\pm 2,\dots$, and $\nu$, just like $\lambda$, is a
certain complex number.

There are four distinct classes of representations and each irreducible
representation is characterized by an eigenvalue of Casimir operator and the
spectrum of the operator $J_3$ \cite{AG,BL,S,S1,S2,W}:
\begin{itemize}
\item {\em Representations unbounded from above and below}, in this case
neither $\lambda + \nu$ nor $\lambda - \nu$ can be integers.

\item {\em Representations bounded below}, with $\lambda + \nu$ being an
integer, and $\lambda - \nu$ not equal to an integer.

\item {\em Representations bounded above,} with $\lambda -  \nu$ being an
integer, and $\lambda + \nu$ not equal to an integer.

\item {\em Representations bounded from above and below,} with $\lambda -
\nu$ and $\lambda + \nu$ both being integers, that yields $\lambda =
k/2, \quad k\in \mathbb Z_{+}$.
\end{itemize}
The nonequivalent representations in each series of irreducible representations
are denoted respectively by $D(\lambda,\nu)$, $D^{+}(\lambda,\nu)$,
$D^{-}(\lambda,\nu)$ and $D(\lambda)$. The representations $D(\lambda,\nu)$,
$D^{+}(\lambda,\nu)$ and $D^{-}(\lambda,\nu)$ are infinite-dimensional;
$D(\lambda)$ is $(2\lambda+1)$-dimensional representation. The irreducible
representations $D^{\pm}(\lambda,\nu)$ and $D(\lambda,\nu)$ are discussed in
details in \cite{AG,BL,S,S1,S2}. Further we restrict ourselves by real
eigenvalues of $J_3$ and the Casimir operator $J^2$.

The infinite-dimensional representations, considered here, are given by linear
unbounded operators in infinite-dimensional linear topological spaces, supplied
with a weak topology. With such a topology a weak convergence (convergence with
respect a functional) - the analog of nonuniform convergence, is associated.
This means that the operators $T(g)$ as elements are continuous on the group,
although indeed not uniformly continuous \cite{S,S1,S2}.

\subsection{Representations unbounded from above and below}

Let $\psi_m$ be eigenstates  of the operators $J_3$ and $J^2$:
\begin{equation}
J_3\psi_m = m \psi_m, \quad J^2\psi_m = \ell(\ell + 1) \psi_m
\label{rep_1}
\end{equation}
where $\ell$ and $m$ are real numbers. Demand that the commutation
relations (\ref{J_01}) and (\ref{J_1}) being satisfied, yields
\begin{eqnarray}
&&J _{-} \psi_m = (\ell +m)\psi_{m-1}, \label{J_1b} \\
&&J _{+}\psi_m  = (\ell - m)\psi_{m+1}, \label{J_2a}.
\end{eqnarray}

Considering the invariance of an inner product $(\psi_m,\psi_{m'})$ with
respect to the infinitesimal rotations generated by $J_i$ we obtain
\begin{align}\label{rot_1}
&(\psi_m,(J_+ - J_-)\psi_{m'}) +(\psi_m(J_+ - J_-),\psi_{m'}) =0,\\
&(\psi_m,\psi_{m'}) =0, \quad m \neq m'.
\end{align}
In particular for $m' = m+1$ one has
\begin{equation}\label{rot_2}
(\psi_m,J_-\psi_{m+1}) -(\psi_m J_+,\psi_{m+1}) =0,
\end{equation}
that implies
\begin{equation}\label{rot_2a}
(l+m +1)(\psi_m,\psi_{m}) - (l -m )(\psi_{m+1},\psi_{m+1}) =0.
\end{equation}
This recursion relationship can be satisfied setting
\begin{equation}\label{pr_1}
(\psi_m,\psi_{m})=\mathcal {N}\,\Gamma(\ell -m +1)\Gamma(\ell + m + 1)
\end{equation}
where $\Gamma$ is the gamma function and $\mathcal {N}$ is an arbitrary
constant. We assume hereafter $\mathcal {N}$ be a positive constant, that does
not restrict generality of consideration.

Setting
\begin{equation}\label{N_1} {\mathcal N}_m = (\mathcal {N}\,\Gamma(\ell
- m +1)\Gamma(\ell + m + 1))^{-\frac{1}{2}},
\end{equation}
we obtain
\begin{equation}\label{pr_1a} \mathcal
(\psi_m,\psi_{m})=|{\mathcal N}_m|^{-2}(-1)^{\sigma(m)}
\end{equation}
where
\begin{eqnarray}
 (-1)^{\sigma(m)}= {\rm sgn}\big(\Gamma(\ell-m +1)\Gamma(\ell+m+1)\big),
\label{sigma}
\end{eqnarray}
and ${\rm sgn}(x)$ is the signum function.

It follows from Eq. (\ref{pr_1a}) that the states ${|\ell,m
\rangle}= {\mathcal N}_m\psi_m $ form the orthonormal basis under
the inner product given by
\begin{equation}\label{nor_1}
\langle m,\ell|\ell,m'\rangle_{\eta} = |{\mathcal N}_m|^2 \eta_{m'
m''}(\psi_m,\psi_{m'})= \delta_{m m'},
\end{equation}
where the indefinite metric is
\begin{align}
\eta_{mm'}= \delta_{mm'}{\rm sgn}\big(\Gamma(\ell-m +1)\Gamma(\ell+m'+1)\big)
\label{metric}
\end{align}

It is easy to see that the operators $J_{\pm}$ act on the states
$|\ell,m\rangle$ as
\begin{align}
& J_{\pm}|\ell,m\rangle = (\ell \mp m)\frac{{\mathcal N}_m}{{\mathcal
N}_{m \pm 1}}|\ell ,m \pm 1\rangle \label{rep_1ab}
\end{align}
From here and Eqs.(\ref{rep_1} - \ref{J_2a}) one obtains
\begin{align}
& J_{+}|\ell,m\rangle = \sqrt{\ell(\ell +1) - m(m +
1)} \;|\ell ,m + 1\rangle  \label{rep_1a}\\
&J_{-}|\ell,m\rangle =  \sqrt{\ell(\ell +1) - m(m - 1)} \;|\ell ,m
-1\rangle \label{rep_1b}\\
&J_{3}|\ell,m\rangle = m|\ell,m\rangle \label{rep_1bc}
\end{align}
One can check that $J_3$ is the self adjoint operator in the indefinite metric
Hilbert space, $J_3= (J_3^\dag)_\eta$, and $ \;J_\pm=(J_{\mp}^\dag)_\eta.$

Now one can start with an arbitrary vector $|\ell,m\rangle$ and apply the
operators $J_{\pm}$ to  obtain any state $|\ell,m'\rangle$. Since the
eigenvalues of $J_3$ can be changed only by multiples of unity, one has $m =
\nu +p$, $p \in \mathbb Z$, where $\nu$ is an arbitrary number with the fixed
value within the given irreducible representation. Thus each irreducible
representation $D({\ell},\nu) $ may be characterized by the given values of two
invariants $\ell$ and $\nu$. In fact the representations $D{(\ell},\nu) $ and
$D({-\ell -1},\nu) $, yielding the same value  $Q= \ell(\ell +1)$ of the
Casimir operator, are equivalent and the inequivalent representations may be
labeled as $D({Q},\nu) $ \cite{W}.

If there exists the number $p_0$ such that $\nu + p_0= \ell$, we have
$J_{+}|\ell,\ell\rangle = 0$ and the representation becomes bounded above. In
the similar manner if for a number $p_1$ one has $\nu + p_1=-\ell $, then
$J_{-}|\ell,-\ell\rangle = 0$ and the representation reduces to the bounded
below. Finally, finite-dimensional unitary representation arises when there
exist possibility of finding  both $J_{+}|\ell,\ell\rangle = 0$ and
$J_{-}|\ell,-\ell\rangle = 0$. It is easy to see that in this case $2\ell,
\;2m$ and $2 \nu$ all must be integers.

Let us consider a realization of the representation $D(\ell,\nu)$ in the
space of analytical functions ${\mathcal F}^{(\ell,\nu)} = \{f(z): z \in
\mathbb C\}$. In this realization the generators $J_{\pm}$ and $J_3$ are
of the forms
\begin{align}
J_{-}= -z^2\partial_{z} + 2\ell z,\; J_{+}=\partial_{z},\; J_{3}=
-z\partial_{z} + \ell,
\end{align}
The functions $f_m^{(\ell,\nu)} (z)=\langle z|\ell,m\rangle$ of the
canonical basis of the representation $D(\ell,\nu)$ take the form
\begin{equation}
\langle z|\ell,m\rangle = {\mathcal N}_m z^{\ell - m}, \label{mon_1a}
\end{equation}
where ${\mathcal N}_m = {(\Gamma(\ell-m+1)\Gamma(\ell +m + 1))}^{-1/2}$ is the
normalization constant. It is easy to see that the eigenvectors
$|\ell,m\rangle$ form the orhonormal basis under the inner product defined as
follows:
\begin{eqnarray}
\langle m,\ell |\ell,m'\rangle_\eta = \overline{\mathcal N}_m{\mathcal
N}_{m''}{\smallsetminus{\hspace{-0.36cm}}\int} \bar{z}^m
\eta_{m'm''}z^{m''}d\mu_{\ell}(z) = \delta_{mm'}
\end{eqnarray}
where $\eta_{m'm''}= (-1)^{\sigma(m')}\delta_{mm'}$ is the
indefinite metric, and
\begin{eqnarray}
d\mu_{\ell}(z) =\frac{\Gamma(2\ell + 2)}{2\pi i}\frac{ d\bar z d z
}{(1+|z|^2)^{2\ell +2}}\label{M_1}
\end{eqnarray}
is the invariant measure.

The inner product of two analytic functions $f(z)$ and $g(z)$ is given by
\begin{eqnarray}\label{pr_2}
\langle f|g\rangle_\eta =  {\smallsetminus{\hspace{-0.36cm}}\int} \bar{f} {\eta
}gd\mu_{\ell}(z).
\end{eqnarray}

Representation $T_g$ acts in space ${\mathcal F}^{(\ell,\nu)}$ as follows
\cite{S1,Vil,and}:
\begin{eqnarray}\label{rep_2}
T_g f(z)=(\bar{\alpha} + \beta z)^{2\ell}f \bigg(\frac{\alpha z -\bar{\beta} }{
\beta z +\bar{\alpha}}\bigg),
\end{eqnarray}
where
\begin{equation}
g = \bigg(\begin{array}{rl}
          \alpha & \;\beta\,\\
          -\bar{\beta} & \;\bar{\alpha}\,
          \end{array}\bigg), \quad |\alpha|^2 + |\beta|^2 = 1
\label{su2}
\end{equation}
Applying (\ref{rep_2}) to the state $|\ell,m\rangle$, we obtain
\begin{align}
&\langle z | T_g|\ell, m\rangle = {\mathcal N}_m(\bar{\alpha} + \beta z)^{\ell
+m} (  \alpha z - \bar{\beta})^{\ell-m}. \label{rep_2a}
\end{align}
It follows from (\ref{rep_2a}) that the representation $D(\ell,\nu)$
is given over the class of analytic functions with the singularities
of the branch-point type and are of orders $\ell + \nu$ and $\ell -
\nu$. Note that in order to define a representation, Mittag-Leffler
uniformization of the functions $f(z)\in{\mathcal F}^{(\ell,\nu)}$
must be carried out \cite{S1}.

The matrix elements ${\mathcal D}^{(\ell,\nu)}_{m m'}(g)$ of the representation
$D(\ell, \nu)$ are defined as the coefficients of the expansion
\begin{align}
T_g|\ell, m'\rangle = \sum {\mathcal D}^{(\ell,\nu)}_{m m'}(g)|\ell .
m\rangle\label{rep_2ab}
\end{align}
The explicit form of the matrix elements in terms of the
hypergeometric function  $F(a,b,c;x)$ is as follows (see Appendix
A):
\begin{align}
{\mathcal D}^{(\ell,\nu)}_{m m'}(g)= &e^{-im\varphi}e^{-im'\psi}(-i)^{m
-m'}z^{(m'-m)/2} (1-z)^{-(m+m')/2} C^{\ell}_{mm'}\nonumber\\
&\times F(-\ell-m,\ell -m +1,m'-m+1;z), \; {\rm if}\; m'> m
\\
{\mathcal D}^{(\ell,\nu)}_{m m'}(g)= &e^{-im\varphi}e^{-im'\psi}(-i)^{m'
-m}z^{(m-m')/2} (1-z)^{-(m+m')/2} C^{\ell}_{m' m}\nonumber \\
&\times F(-\ell-m',\ell -m' +1, m-m'+1;z), \;{\rm if}\; m' < m
\end{align}

where $z=\sin^2(\theta/2)$, and
\begin{align}
C^{\ell}_{mm'}=\frac{1}{(m'-m)!}\bigg(\frac{\Gamma(\ell - m +1)\Gamma(\ell + m'
+ 1)}{\Gamma(\ell + m +1)\Gamma(\ell - m' + 1)}\bigg)^{\frac{1}{2}}.
\end{align}

Since the spectrum of the operator $J_3$ has the form $m= p+ \nu, \; p=0,
\pm1,\pm2, \dots$, the matrix elements ${\mathcal D}^{(\ell,\nu)}_{m m'}(g)$
are $s$-valued functions over the group if $\nu$ is a rational number, $\nu =
r/s$, and are infinity-valued if $\nu$ is irrational number.

Thus, the representation $D(\ell, \nu)$ is multiple-valued infinite-dimensional
representation of the rotation group. It is given in an infinite-dimensional
space, in which the convergence is of the weakest type. The elements $
{\mathcal D}^{(\ell,\nu)} \in D(\ell, \nu)$ of such a space are the generalized
functions, which can be considered as linear functionals on some space of basic
functions. Such representations are exact representations of an
infinity-sheeted universal covering group $\widetilde{SU(2)}$ of the rotation
group (for details see \cite{S,S1,S2}).

\subsection{Representations bounded above}

This representation is characterized by the eigenvalue $\ell$ of the
highest-weight state:  $|\ell,0\rangle$ such that $J_{+}|\ell,0\rangle = 0$ and
$ J_{3} |\ell,0\rangle = \ell|\ell,0\rangle$. It can be obtained formally from
the representation unbounded from above and below setting $m =\ell -n$, $n=
0,1,2, \dots$. It is convenient to consider the orthonormal states
$|\ell,n\rangle$  instead of $|\ell,m\rangle$. The eigenvectors
$|\ell,n\rangle$ form a basis in the space of the representation
$D^{-}({\ell,\nu})$, where the operator $J_3$ acts as follows:
\begin{equation}
J_3|\ell,n\rangle=(\ell-n)|\ell,n\rangle, \quad n=0,1,\dots
\end{equation}
The action of the operators $\{J_{\pm}\}$ on the states is given by
\begin{align*}
 J _{+} |\ell,n\rangle =
\sqrt{n(2\ell-n+1))}|\ell,n-1\rangle \\
J _{-} |\ell,n\rangle = \sqrt{(n+1) (2\ell -n)}|\ell,n+1\rangle
\end{align*}

We consider a suitable realization of the representation $D^{-}({\ell,\nu})$ in
the space of entire analytical functions ${\mathcal F}^{\ell} = \{f(z): z \in
\mathbb C\}$. In this realization the generators $J_{\pm}$ and $J_3$ act as the
first order differential operators:
\begin{align}
J_{-}= -z^2\partial_{z} + 2\ell z,\; J_{+}=\partial_{z},\; J_{3}=
-z\partial_{z} + \ell,
\end{align}
The monomials
\begin{equation}
\langle z|\ell,n\rangle = {\mathcal N}_n z^n, \label{mon_1}
\end{equation}
where ${\mathcal N}_n = {(\Gamma(n+1)\Gamma(2\ell - n + 1))}^{-1/2}$ is
the normalization constant, form an orthogonal basis for holomorphic
functions analytical in $\mathbb C$, and satisfy
\begin{align}
&(z^n,z^p): =\frac{\Gamma(2\ell + 2)}{2\pi
i}{\smallsetminus{\hspace{-0.36cm}}\int} \frac{ \bar{z}^n z^p d\bar z
d z }{(1+|z|^2)^{2\ell +2}} \nonumber \\
&= \Gamma(n+1)\Gamma(2\ell - n + 1)\delta_{np}. \label{p1b}
\end{align}
For $n > 2\ell$ the value of r.h.s. is given by the analytical continuation of
the gamma function \cite{S}.

It follows from Eq.(\ref{p1b}) that the states $|\ell,n\rangle$ form the
orhonormal basis under the indefinite metric inner product defined as
follows:
\begin{eqnarray}
\langle n,\ell |\ell,p\rangle_\eta = \eta_{pp'}(\langle z|\ell,n\rangle,\langle
z|\ell,p'\rangle) = \delta_{np},
\end{eqnarray}
where $\eta_{np}= (-1)^{\sigma(n)}\delta_{np}$ and
\begin{align*}
&(-1)^{\sigma(n)}=\left\{\begin{array}{l} 1,\; {\rm if}\; 2\ell -n >0\\
(-1)^{n+1}\,{\rm sgn}(\sin2\pi\ell),\; {\rm if}\; n -2\ell >0
  \end{array}\right.
\end{align*}

An arbitrary state of the representation is an entire function of the type
\begin{equation}\label{f_1}
f(z) = \sum^{\infty} _{n=0}f_n \langle z|\ell,n\rangle.
\end{equation}
The inner product of two entire functions $f(z)$ and $g(z)$ is constructed
as follows:
\begin{eqnarray}\label{pr_21}
\langle f|g\rangle_\eta =  \frac{\Gamma(2\ell +2)}{2\pi
i}{\smallsetminus{\hspace{-0.36cm}}\int}_D \frac{ \bar{f} {\eta }g\,d\bar z d
z}{(1+|z|^2)^{2+2\ell}}
\end{eqnarray}

\subsection{Representations bounded below}

Setting $m=n - \ell$, for the representation bounded below  we have
\begin{equation}
J_3|\ell,n\rangle=(n -\ell)|\ell,n\rangle, \quad n=0,1,\dots
\end{equation}
The representation is characterized by the eigenvalue $\ell$ of the
highest-weight state:  $|\ell,0\rangle$ such that $J_{-}|\ell,0\rangle = 0$ and
$ J_{3} |\ell,0\rangle = -\ell|\ell,0\rangle$.

The action of the operators $\{J_{\pm}\}$ on the state $|\ell,n\rangle$ is
given by
\begin{align*}
 J _{-} |\ell,n\rangle =
\sqrt{n(2\ell-n+1))}|\ell,n-1\rangle \\
J _{+} |\ell,n\rangle = \sqrt{(n+1) (2\ell -n)}|\ell,n+1\rangle .
\end{align*}

We consider a  realization of the representation $D^{+}({\ell,\nu})$ in the
space of analytical functions ${\mathcal F}^{\ell} = \{f(z): z\in \mathbb C\}$,
such that $z^{-2\ell}f(z)$ is the meromorphic function. In this realization the
generators $J_{\pm}$ and $J_3$ act as the following differential operators:
\begin{align}
J_{-}= -z^2\partial_{z} + 2\ell z,\; J_{+}=\partial_{z},\; J_{3}=
-z\partial_{z} + \ell,
\end{align}
The monomials
\begin{equation}
\langle z|\ell,n\rangle = {\mathcal N}_n z^{2\ell- n}, \\
\end{equation}
where ${\mathcal N}_n = {(n!\Gamma(2\ell - n + 1))}^{-1/2}$ is the  same
normalization constant as in (\ref{mon_1}), form the orthonormal basis such
that
\begin{eqnarray*}
\langle n,\ell |\ell,p\rangle_\eta =\frac{(-1)^{\sigma(n)}\Gamma(2\ell +
2)}{2\pi i\, n!|\Gamma(2\ell - n + 1)|}{\smallsetminus{\hspace{-0.36cm}}\int}
\frac{ \bar{z}^{2\ell-n} z^{2\ell-p}d\bar z d z}{(1+|z|^2)^{2\ell+2}}=
\delta_{np},
\end{eqnarray*}
$\eta_{np}=(-1)^{\sigma(n)}\delta_{np}$ being the indefinite metric, and
\begin{align*}
&(-1)^{\sigma(n)}=\left\{\begin{array}{l} 1,\; {\rm if}\; 2\ell -n >0,\\
(-1)^{n+1}\,{\rm sgn}(\sin2\pi\ell),\; {\rm if}\; n -2\ell >0.
  \end{array}\right.
\end{align*}

An arbitrary state of the representation is a function of the type
\begin{equation}\label{f_1a}
f(z) = \sum^{\infty} _{n=0}f_n \langle z|\ell,n\rangle.
\end{equation}
The inner product of the functions $f(z)$ and $g(z)$ is constructed as
above (see Eq.(\ref{pr_21})):
\begin{eqnarray}\label{p2e}
\langle f|g\rangle_\eta =  \frac{\Gamma(2\ell +2)}{2\pi
i}{\smallsetminus{\hspace{-0.36cm}}\int} \frac{ \bar{f} {\eta }g d\bar z d
z}{(1+|z|^2 )^{2+2\ell}}.
\end{eqnarray}

\section{Infinite-dimensional representations and Dirac monopole problem}

 As well known any choice of the vector potential $\mathbf A$
being compatible with a magnetic field ${\mathbf B}$ of Dirac monopole
must have  singularities (the so-called strings), and one can write
\[
{\mathbf B}={\rm rot}{\mathbf A} + {\mathbf h},
\]
where ${\mathbf h}$ is the magnetic field of the string.

There is an ambiguity in the definition of the vector potential. For
instance, Dirac introduced the vector potential as \cite{Dir}
\begin{equation}
{\mathbf A}_{\mathbf n}= q\frac{{\mathbf r}\times {\mathbf n}}
{r(r - {\mathbf n} \cdot{\mathbf r})} \label{d_str}
\end{equation}
where the unit vector $\mathbf n$ determines the direction of string
$S_{\mathbf n}$, passing from the origin of coordinates to $\infty$ , and
\begin{equation}
{\mathbf h}_{\mathbf n}  = 4\pi q{\mathbf n}\int _{0}^\infty
\delta^3(\mathbf r - \mathbf n \tau) d \tau.
\end{equation}
Schwinger's choice is \cite{Sw_1}:
\begin{equation}
{\mathbf A^{SW}}= \frac{1}{2}\bigl({\mathbf A}_{\mathbf n}+
{\mathbf A}_{-\mathbf n} \bigr), \label{sw}
\end{equation}
with the string being propagated from $-\infty$ to $\infty$. Both
vector potentials yield the same magnetic monopole field, however
the quantization is different. The Dirac condition is $2\mu=p$,
while the Schwinger one is $\mu=p, \; p\in \mathbb Z$.

These two strings are members of a family $\{{S}^{\kappa}_{\mathbf
n} \}$ of  {\it weighted strings}, which magnetic field is given by
\footnote{Previously \cite{NF,NF1}, we have used the other
definition of the string weight, namely, ${\mathbf
h}^{\kappa}_{\mathbf n}= \kappa{\mathbf h}_{\mathbf n} + (1-
\kappa){\mathbf h}_{-\mathbf n}$. To compare them, one should make
substitution $\kappa \rightarrow (1+\kappa)/2$.}
\begin{align}
&{\mathbf h}^{\kappa}_{\mathbf n}= \frac{1+ \kappa}{2}{\mathbf
h}_{\mathbf n} + \frac{1- \kappa}{2}{\mathbf h}_{-\mathbf n}
\label{str}
\end{align}
where $\kappa$ is the weight of a semi-infinite Dirac string. The
respective vector potential reads
\begin{align}
{\mathbf A}^\kappa_{\mathbf n} = \frac{1+ \kappa}{2}{\mathbf
A}_{\mathbf n} + \frac{1- \kappa}{2}{\mathbf A}_{-\mathbf n},
\label{A_1}
\end{align}
and since ${\mathbf A}^\kappa_{-\mathbf n}={\mathbf
A}^{-\kappa}_{\mathbf n}$, we obtain the following equivalence
relation: ${S}^{\kappa}_{-\mathbf n} \simeq{S}^{-\kappa}_{\mathbf n}
$. This implies, that the string ${S}^{\kappa}_{\mathbf n}$ is
invariant under the following discrete transformation:
\begin{align}
\kappa \rightarrow -\kappa, \;q \rightarrow -q, \;\mathbf r
\rightarrow -\mathbf r .\label{A1a}
\end{align}

Note that two arbitrary strings $S^\kappa_{\mathbf n}$ and
$S^\kappa_{\mathbf n'}$ are related by
\begin{eqnarray}
A^{\kappa'}_{\mathbf n'}= A^\kappa_{\mathbf n}+ d\chi. \label{ag2a}
\end{eqnarray}
and vice versa. Besides, an arbitrary transformation of the strings
$S^\kappa_{\mathbf n} \rightarrow S^{\kappa'}_{\mathbf n'}$ can be
realized as combination $S^\kappa_{\mathbf n} \rightarrow
S^{\kappa}_{\mathbf n'}$ and $S^\kappa_{\mathbf n} \rightarrow
S^{\kappa'}_{\mathbf n}$, where the first transformation preserving
the weight of the string is rotation, and the second one results in
changing of the weight string $\kappa \rightarrow \kappa'$ without
changing its orientation \cite{NF,NF1}.

Let denote by $\mathbf n'= g\mathbf n , g\in\rm SO(3)$, the left
action of the rotation group induced by $S^\kappa_{\mathbf n}
\rightarrow S^\kappa_{\mathbf n'}$. From rotational symmetry of the
theory it follows this gauge transformation can be undone by
rotation $\mathbf r \rightarrow  \mathbf r g$ as follows
\cite{Sw_1,BP,NF}:
\begin{align}
&A^{\kappa}_{\mathbf n'}(\mathbf r)= A^{\kappa}_{\mathbf n}(\mathbf r')=
A^\kappa_{\mathbf n}(\mathbf r)+
d\alpha({\mathbf r}; g) , \label{g_0}\\
&\alpha(\mathbf r;g)=  \int_{\mathbf r}^{\mathbf r'} \mathbf
A^\kappa_{\mathbf n}(\boldsymbol \xi) \cdot d \boldsymbol \xi, \quad
\mathbf r' =  \mathbf r g \label{g_1c}
\end{align}
where the integration is performed along the geodesic
$\widehat{\mathbf  r \,\mathbf r'}\subset S^2$. This gauge
transformation may be written also as
\begin{align}
A^{\kappa}_{\mathbf n'}(\mathbf r)= A^\kappa_{\mathbf n}(\mathbf r)-
d\Omega (\mathbf n ,\mathbf n';\mathbf r ),
\end{align}
where $\Omega (\mathbf n ,\mathbf n';\mathbf r )$ is the solid angle
of the geodesic simplex with the vertices $(\mathbf n ,\mathbf
n';\mathbf r )$ \cite{Wu2,Wu3,FH}.

Now returning to the transformation $S^\kappa_{\mathbf n} \rightarrow
S^{\kappa'}_{\mathbf n}$ we obtain
\begin{eqnarray}
&&A^{\kappa'}_{\mathbf n} = A^{\kappa}_{\mathbf n} -
d\chi_{\mathbf n}, \nonumber \\
 &&d\chi_{\mathbf n} = q{(\kappa' - \kappa)}\frac{(\mathbf r \times
\mathbf n)\cdot d\mathbf r}{r^2- (\mathbf n \cdot \mathbf r)^2},
\label{A_02}
\end{eqnarray}
$\chi_{\mathbf n}$ being polar angle in the plane orthogonal to
${\mathbf n}$. In particular, if $\kappa' = -\kappa$ we obtain the
mirror string: $S^\kappa_{\mathbf n} \rightarrow
S^{\kappa}_{-\mathbf n}\simeq S^{-\kappa}_{\mathbf n} $.

For a non relativistic charged particle in the field of a magnetic
monopole the total angular momentum
\begin{equation}
{\mathbf J} = {\mathbf r} \times \left({\mathbf p} - e{\mathbf
A}\right) - \mu\frac{\mathbf r}{r} \label{eq1c}
\end{equation}
having the same properties as a standard angular momentum, obeys the
following commutation relations:
\begin{eqnarray}
&&[H, {\mathbf J}^2] = 0, \quad [H, J_i] = 0,\quad  [{\mathbf J}^2,
J_i] =
0, \label{eq5a} \\
&&[J_i, J_j] = i\epsilon_{ijk}J_k \label{eq5}
\end{eqnarray}
where $H$ is the Hamiltonian.

These commutation relations fail on the string \cite{Zw_1, Str}.
However, it is found that $H$ and $\mathbf J$ may be extended to
self-adjoint operator satisfying the commutation relations of the
rotation group, with $H$ an invariant, namely
\[
[H, J_i] = 0,\quad [J_i, J_j] = i\epsilon_{ijk}J_k ,
\]
and this is true for any value of $\mu$. However, requiring that the $J_i$
generate a finite dimensional representation of the rotation group and not just
the Lie algebra, one obtains that $\mu$ must be quantized and only the values
$2\mu = 0,\pm 1,\pm 2, \dots $ are allowed \cite{H}.

Taking into account the spherical symmetry of the system, the vector
potential can be considered as living on the two-dimensional sphere of the
given radius $r$ and being taken as \cite{Wu1,Wu2}
\begin{eqnarray}
{\mathbf A_N} =  q\frac{1-\cos{\theta}}{r\sin{\theta}}\;
\hat{\mathbf e}_{\varphi}, \quad {\mathbf A_S} =
-q\frac{1+\cos\theta}{r\sin\theta}\; \hat{\mathbf e}_{\varphi}
\label{eq1b}
\end{eqnarray}
where $(r,\theta,\varphi)$ are the spherical coordinates, and while
${\mathbf A_{N}}$ has singularity on the south pole of the sphere,
${\mathbf A_{S}}$ is singular on the north one. In the overlap of
the neighborhoods  covering the sphere $S^2$ the potentials
${\mathbf A_N}$ and ${\mathbf A_S}$  are related by the following
gauge transformation:
\[
A_S = A_N - 2qd\varphi.
\]
This is the particular case of (\ref{A_02}), when $\kappa = 0$ and
$\kappa' =1$.

Choosing the vector potential as $\mathbf A_N$ we have
\begin{align}
J_{\pm}= &\;e^{\pm i\varphi}\bigg(\pm\frac{\partial}{\partial\theta}
+i\cot\theta \frac{\partial}{\partial\varphi} -
\frac{\mu\sin\theta}{1+\cos\theta} \bigg)\\
J_3=&-i\frac{\partial}{\partial\varphi} - \mu\\
{\mathbf J^2} =&-\frac{1}{\sin{\theta}}
\frac{\partial~}{\partial\theta}\left(\sin{\theta}
\frac{\partial~}{\partial\theta}\right) -
\frac{1}{\sin^2{\theta}}\frac{\partial^2~}{\partial\varphi^2}  \nonumber\\
&+\frac{2i\mu}{1 +\cos{\theta}}\frac{\partial~}{\partial\varphi}
+\mu^2\frac{1 - \cos{\theta}}{1 + \cos{\theta}} +\mu^2 \label{eq7}
\end{align}
where  $J_{\pm} = J_1 \pm iJ_2$ are the raising and the lowering operators for
$J_3$.

Now starting with Schr\"odinger's equation
\begin{equation}
\hat H\Psi = E \Psi, \label{eq01}
\end{equation}
and introducing $\Psi= R(r)Y(\theta,\varphi)$, we get for the
angular part the following equation:
\begin{eqnarray}
&&{\mathbf J^2}Y(\theta,\varphi) = \ell({\ell} + 1)Y(\theta,
\varphi). \label{eq7a}
\end{eqnarray}

By substituting
\begin{eqnarray*}
Y =e^{i(m+\mu)\varphi}z^{(m +\mu)/2}(1-z)^{(m - \mu)/2}F(z),
\end{eqnarray*}
into Eq.(\ref{eq7a}), where $z=(1-\cos\theta)/2$ and $m$ is an
eigenvalue of $J_3$, we find that $F(z)$ is a solution of the
hypergeometric equation:
\begin{align}
&z(1-z)\frac{d^2F}{dz^2} +\bigl(c-(a+b+1)z\bigr)\frac{dF}{dz}-abF=0
\label{hyp1} \\
&a = m - \ell, \; b = m + \ell + 1, \; c = 1 + m + \mu, \label{sols}
\end{align}

\subsection{Representation bounded above or below}

As is known the hypergeometric function $F(a,b;c;z)$ reduces to a
polynomial of degree $n$ in $z$ when $a$ or $b$ is equal to $-n,
\;(n = 0,1,2, \dots)$, and the respective  solution of
Eq.(\ref{hyp1}) is of the form \cite{abr,and}
\begin{eqnarray}
F=z^{\rho}{(1-z)}^{\sigma} p_n(z) \label{pol}
\end{eqnarray}
where $p_n(z)$ is a polynomial in $z$ of degree $n$. Here we are looking
for the solutions, like this of the Schr\"odinger equation (\ref{eq7a}).

The requirement of the wave function being single valued force us to
take $\alpha = m +\mu$ as an integer and general solution is given
by
\begin{equation}\label{gsol_1}
 Y_{\ell}^{(\mu,n)}=e^{i\alpha\varphi}Y_n^{(\delta,\gamma)}(u),
\end{equation}
where $u = \cos\theta$, and
\begin{equation}\label{pol_2}
Y_n^{(\delta,\gamma)}(u) =C_n\,(1-u)^{\delta/2}(1+u)^{\gamma/2}
P_n^{(\delta,\gamma)}(u),
\end{equation}
$P_n^{(\delta,\gamma)}(u)$ being the Jacobi polynomials, and the normalization
constant $C_n$ is given by
\begin{equation*}
C_n=\Bigg(\frac{2 \pi \, 2^{\delta +\gamma +1}\Gamma(n +\delta +1) \Gamma(n
+\gamma +1)}{(2n +\delta +\gamma +1)\Gamma(n+1)\Gamma(n +\delta
+\gamma+1)}\Bigg)^{-1/2}
\end{equation*}

There are four distinct classes of solutions and each solution is
characterized by an eigenvalue of Casimir operator and the
eigenvalue spectrum of $J_3$:
\begin{align*}
&\stackrel{{\quad}-\;(\mu,n)}{Y_{\ell\pm \mu}}
=e^{i\alpha\varphi}Y_n^{(\alpha,\beta)}\left\{\begin{array}{l}
m= \ell -n,\;\ell+\mu \in {\mathbb Z}_{+}\\
m= -\ell- n-1, \; \ell-\mu \in {\mathbb Z}_{+}
  \end{array}\right .\\
&\stackrel{{\quad}+\;(\mu,n)}{Y_{\ell\mp \mu}}
=e^{i\alpha\varphi}Y_n^{(-\alpha,-\beta)}\left\{\begin{array}{l} m=
n-\ell,\;\ell-\mu \in {\mathbb Z}_{+}\\
m= \ell+ n+1, \; \ell+\mu \in {\mathbb Z}_{+}
  \end{array}\right .
\end {align*}
where we set $\alpha = m+\mu$ and $\beta = m-\mu$.

The obtained solutions belong to  the indefinite metric Hilbert
space ${\cal H}^\eta$, with the metric being
\begin{align*}
&\eta_{np}= \left\{\begin{array}{l}
\delta_{np},\; {\rm if}\; 2\ell -n >0\\
\delta_{np}(-1)^{n+1}\,{\rm sgn}(\sin2\pi\ell),\; {\rm if}\; n
-2\ell >0.
  \end{array}\right.
\end{align*}

The functions $\stackrel{{\quad}-\;(\mu,n)}{Y_{\ell \pm \mu}}$ form the basis
of the infinite-dimensional representations bounded above, being denoted
respectively by $D^-(\ell,-\mu)$ and $\widetilde D^-(\ell,-\mu)$ , and the
functions $\stackrel{{\quad}+\;(\mu,n)}{Y_{\ell \mp \mu}}$ form the basis of
the representations $D^{+}(\ell,-\mu)$ and $\widetilde D^{+}(\ell,-\mu)$
bounded below.

A similar consideration can be done for the vector potential
$\mathbf A_S$. In this case $\beta =m-\mu\in \mathbb Z$ and the
corresponding wave functions
\begin{align*}
&\stackrel{{\quad}-(-\mu,n)}{Y_{\ell\mp \mu}}
=e^{i\beta\varphi}Y_n^{(\alpha,\beta)}\left\{\begin{array}{l} m=
\ell -n,\;\ell-\mu \in {\mathbb Z}_{+}\\
m= -\ell- n-1, \; \ell+\mu \in {\mathbb Z}_{+}
  \end{array}\right .\\
&\stackrel{{\quad}+(-\mu,n)}{Y_{\ell\pm \mu}}
=e^{i\beta\varphi}Y_n^{(-\alpha,-\beta)}\left\{\begin{array}{l}
m= n-\ell,\;\ell+\mu \in {\mathbb Z}_{+} \\
m= \ell+ n+1, \; \ell-\mu \in {\mathbb Z}_{+}
  \end{array}\right .
\end{align*}
form a complete set of orthonormal basis of the infinite-dimensional
representation $D^{\pm}(\ell ,\mu)$ and $\widetilde D^{\pm}(\ell
,\mu)$.

Thus, we find the following series of representations:
\begin{align*}
&\ell-\mu \in {\mathbb Z}_{+} \Rightarrow\left\{\begin{array}{l}
D^+(\ell ,-\mu): \; m= n - \ell\\
D^-(\ell ,\mu): \;m=\ell - n \\
\widetilde D^+(\ell,\mu): \;m= n +\ell +1\\
\widetilde D^-(\ell,-\mu): \; m= -\ell -n -1\\
 \end{array}\right . \\
&\ell+\mu \in {\mathbb Z}_{+} \Rightarrow\left\{\begin{array}{l}
D^+(\ell,\mu): \;m= n-\ell\\
D^-(\ell ,-\mu): \; m= \ell -n\\
\widetilde D^+(\ell ,-\mu): \; m= n + \ell +1\\
\widetilde D^-(\ell ,+\mu): \;m= -\ell- n-1\\
 \end{array}\right .
\end{align*}
where $n = 0,1,2,\dots$. Notice, that while the representations
$D^\pm(\ell,\pm\mu)$ and $D^\pm(\ell,\mp\mu)$ are irreducible, the
representations $\widetilde D^\pm(\ell,\pm\mu)$ and $\widetilde
D^\pm(\ell,\mp\mu)$ are partially reducible \cite{W}.

Taking into account the following restriction:  $\ell(\ell+1)-\mu^2
\geq 0$, emerging from the Schr\"odinger equation, the allowed
values of $\ell$ are found to be
\begin{align*}
&\ell+ \mu \in {\mathbb Z}_{+}\,\Rightarrow \, \ell= -\mu + [2\mu] +k, \quad k = 0,1,2,\dots \\
&\ell- \mu \in {\mathbb Z}_{+}\,\Rightarrow \, \ell= \mu + k, \quad k = 0,1,2,\dots
\end{align*}
where $[2\mu]$ denotes the integer part of $2\mu$. Thus, for $\mu$
being arbitrary valued, the representations corresponding to
$\ell+\mu \in {\mathbb Z}_{+}$ and $\ell-\mu \in {\mathbb Z}_{+}$
are not equivalent. However, if $2\mu \in \mathbb Z$, then $\ell+
\mu$ and $\ell- \mu$  both are integers, and the representations has
been obtained above become finite-dimensional equivalent
representations, and vice versa. Thus, we see that the Dirac
quantization rule is related to the finite-dimensional representations
of the rotation group. \\

\subsection{Representation unbounded from above and below}

Let us start with a generic case of the weighted string $S^\kappa_{\bf n}$
crossing the sphere at the north and south poles. We assume $\mathbf
n=(0,0,1)$. With this choice the vector potential takes the form:
\begin{equation}
{\mathbf A}^{\kappa} =\frac{1+\kappa}{2}{\mathbf A}_{S} +
\frac{1-\kappa}{2}{\mathbf A}_{N} \label{eq15}
\end{equation}
The corresponding solution of Eq. (\ref{eq7a}) is given by (see Appendix B)
\begin{align}
Y\propto {e}^{-i\kappa\mu\varphi} {e}^{im\varphi}
z^{(m+\mu)/2}(1-z)^{(m-\mu)/2}Y_{\kappa}(z),\label{eq10}
\end{align}
where
\begin{align}
Y_{\kappa}(z)=& \frac{1+\kappa}{2}F(m-\ell, m+\ell+1,1+ m -\mu; 1-z) \nonumber\\
&+\frac{1-\kappa}{2}F(m-\ell, m+\ell+1,1+ m +\mu; z) \label{eq9}
\end{align}
Here $m$ is an eigenvalue of the operator $J_3$, and  its spectrum
is of the form:
\begin{equation}\label{Spec1} m  =n+ \nu, \; n=0,\pm 1,\pm 2,
\dots.,
\end{equation}
$\nu$ being  an arbitrary real number.

The set of wave functions $\left\{Y_{\kappa,\ell}^{(\mu,m)}\right\}$ such that
\begin{align}
&Y_{\kappa,\ell}^{(\mu,m)}= C^{\ell}_{\mu m}{e}^{-i\kappa\mu\varphi}
{e}^{im\varphi}z^{(m+\mu)/2}(1-z)^{(m-\mu)/2}Y_{\kappa}(z),\nonumber \\
&m  =n + \nu, \; n=0,\pm 1,\pm 2, \dots,\label{rep_4}
\end{align}
$C^{\ell}_{\mu m}$ being a normalization constant, form the complete
orthonormal canonical basis of the representation $D^\kappa(\ell,\nu)$ in the
indefinite-metric Hilbert space ${\cal H}^\eta$. The indefinite metric is given
by
\begin{equation}\label{eq8a}
 \eta_{m m'} = (-1)^{\sigma(m)}\delta_{m m'},
\end{equation}
where $(-1)^{\sigma(m)}= {\rm sgn}\big(\Gamma(\ell-m +1)\Gamma(\ell+m+1)\big)$,
and there is no any restriction on $\mu$.

Since the set of functions $\{Y_{\kappa,\ell}^{(\mu,m)} \}$ forms the
orthonormal basis in the indefinite metric Hilbert space ${\cal H}^\eta$ of the
irreducible infinite-dimensional representation $D^\kappa(\ell,\nu)$, any
solution of the Schr\"odinger equation (\ref{eq7a}) can be expanded as
\begin{equation}
\Psi = \sum_{l,m} C_{\ell ,m}Y_{\kappa,\ell}^{(\mu,m) } \label{eq3d}
\end{equation}
where $\mu$ is an {\em arbitrary parameter}.

The obtained infinite-dimensional representation
$D^{\kappa}(\ell,\nu) $ being in general case multi-valued, depends
on arbitrary parameters $\nu$ and $\kappa$, and hence, the class of
all representations is too large. Further simplification can be done
by relating in $\nu$ with the weight $ \kappa$ of the string in such
a way, that in particular cases of $\kappa =\pm 1$ and $\kappa =0$
the of functions $\{Y_{\kappa,\ell}^{(\mu,m)} \}$ being associated
with the Dirac and Schwinger string, respectively, behave in
appropriate way at the north and south poles of the sphere (for
details see Appendix B). This leads to the following condition: $\nu
= \kappa\mu$, that provides also the functions
$Y_{\kappa,\ell}^{(\mu,m)}$ of the canonical basis be single-valued.
Then the spectrum of the operator $J_3$ takes the form:
\begin{equation}\label{Spec1a} m  =n + \kappa\mu, \; n=0,\pm 1,\pm 2,
\dots.
\end{equation}

Denoting the representation by $D(\ell,\kappa\mu)$ we find that the
representation $D(\ell,\kappa\mu)$ becomes the representation
$D^{+}(\ell,\kappa\mu)$ bounded below, if $\ell + \kappa\mu$ is an integer, and
$\ell  -\kappa\mu$ is not equal to an integer. In a similar way,
$D(\ell,\kappa\mu)$ reduces to the representation $D^{-}(\ell,\kappa\mu)$
bounded above, with $\ell - \kappa\mu$ being an integer, and $\ell  +
\kappa\mu$ not equal to an integer.

Let us consider some particular cases related to the Dirac and Schwinger
strings, starting with $\kappa=-1$. The corresponding solution can be written
as follows (for details see Appendix B):
\begin{align}
&Y^{(\mu,n)}_{-1,\ell} \propto e^{i
n\varphi}z^{n/2}(1-z)^{n/2-\mu}F(a,b,c;z), \nonumber\\
&a = n - \mu - \ell, \; b =n - \mu  + \ell + 1, \; c =1+n, \nonumber
\end{align}
and the function $ Y^{(\mu,n)}_{-1,\ell}$ being regular at the point $z=0$, has
singularity at the south pole ($z=1$), where the Dirac string $S_{\mathbf
n}^{-1}$ crosses the sphere.

The choice of $\kappa = 1$ corresponds to the string $S_{\mathbf n}^{1}$
crossing the sphere at the north pole ($z=0$), and the solution of the equation
(\ref{eq7a}) being regular at the point $z=1$ is given by
\begin{align}
&Y^{(\mu,n)}_{1,\ell} \propto e^{i
n\varphi}z^{n/2+\mu}(1-z)^{n/2}F(a,b,c;1-z),\nonumber\\
&a = n + \mu - \ell, \; b =n + \mu  + \ell + 1, \; c =1+n.  \nonumber
\end{align}

Finally, setting $\kappa =0$ we have the Schwinger case:
\begin{align}
Y^{(\mu,n)}_{0,\ell} \propto& \;{e}^{i
n\varphi}z^{(n+\mu)/2}(1-z)^{(n-\mu)/2}\big ( F(n-\ell, n +\ell+1,1+ n -\mu;
1-z) \nonumber\\
&+ F(n-\ell, n+\ell+1,1+ n +\mu; z)\big ),
\end{align}
with the string $S_{\mathbf n}^{0}$ being propagated from $-\infty$ to
$\infty$.

Later on the strings $S_{\mathbf n}^{-1}$, $S_{\mathbf n}^{1}$ and $S_{\mathbf
n}^{0}$ we will call the {\em fundamental strings}. The related representations
are $D(\ell,\mu), \;D(\ell,-\mu)$ and $D(\ell,0)$, and for $\ell -\mu$ or $\ell
+ \mu$ being integer, one obtains the representations bounded below or above:
$D^\pm(\ell,\pm\mu)$ and $D^\pm(\ell,\mp\mu)$. The respective solutions we will
call {\em generalized monopole harmonics}. When $n+\alpha$, $n+\beta$ and
$n+\alpha+\beta$ all are integers $\geq 0$, the generalized monopole harmonics
are reduced to the {\em monopole harmonics} introduced by Wu and Yang
\cite{Wu2}. The imposed here restrictions on the values of $n,\alpha$ and
$\beta$ yield the finite-dimensional unitary representation of the rotation
group and the Dirac quantization condition.

Note, that the representation $D(\ell,0)$ may be realized only as
infinite-dimensional representation unbounded from above and below, or finite
dimensional representation when $\ell \in \mathbb Z$. It is easy to see that in
this case $\mu$ has to be an integer, that implies Schwinger's quantization of
the magnetic charge.

Returning to the general case, let us consider two strings $S_{\mathbf
n}^\kappa$ and $S_{\mathbf n}^{\kappa'}$.  The corresponding vector potentials
are ${\mathbf A}^{\kappa}$ and ${\mathbf A}^{\kappa'}$ (see Eq.(\ref{eq15})),
and the computation yields
\begin{equation}\label{w1}
 e(A^{\kappa} -  A^{\kappa'}) = (\kappa' - \kappa)\mu d\varphi.
\end{equation}
Recalling that ${\mathbf A}^{\kappa}$ and ${\mathbf A}^{\kappa'}$ are connected
by a gauge transformation:
\begin{equation}\label{w2}
 ie({\mathbf A}^{\kappa'}-{\mathbf A}^{\kappa}) = { e}^{-i\chi}\nabla{\rm
 e}^{i\chi},
\end{equation}
we find that the gauge transformation $\exp(i\chi)$ must satisfy
\[
i\mu(\kappa - \kappa') = \frac{d}{d\varphi}(\ln{e}^{i\chi})
\]
The solution of this equation is $\chi = \mu(\kappa - \kappa')\varphi$. If $p$
is a winding number of the map $\exp\big(i\chi(\varphi)\big):\; S^1 \rightarrow
U(1) $, then
\begin{equation}\label{w3}
 \mu(\kappa - \kappa') = p, \quad p \in \mathbb Z
\end{equation}
that provides $\mu(\kappa - \kappa')$ be an integer. In particular, choosing
$\kappa' = \pm 1$ we obtain $\kappa\mu = \pm \mu + p$, and if $\kappa' = 0 $
one has $\kappa\mu = p$, $p \in \mathbb Z$.

We say that the strings $S_{\mathbf n}^\kappa$ and $S_{\mathbf n}^{\kappa'}$
are {\em gauge equivalent}, $S_{\mathbf n}^\kappa\simeq S_{\mathbf
n}^{\kappa'}$, if $\kappa \mu = \kappa'\mu \mod\mathbb Z$. As we can easily
see, the fundamental strings induce the following classes of gauge equivalent
strings:
\begin{align*}
&S^\kappa_{\mathbf n}\simeq S^1_{\mathbf n}:\quad \kappa \mu = \mu  +p\\
&S^{\kappa}_{\mathbf n} \simeq S^{-1}_{\mathbf n}\simeq S^1_{-\mathbf n}:\quad \kappa \mu =- \mu  +p \\
&S^\kappa_{\mathbf n}\simeq S^0_{\mathbf n}:\quad \kappa \mu =p,\;p=0,\pm 1,\pm
2, \dots .
\end{align*}
The related classes of irreducible representations are $\{D(\ell,\mu)\}$,
$\{D(\ell,-\mu)\}$ and $\{D(\ell,0)\}$, respectively, and the spectrum of the
operator $J_3$ is found to be:
\begin{align*}
&D(\ell,\mu): \;m  =n +\mu, \; {\rm if}\; (\kappa-1)\mu \in \mathbb Z ,\\
&D(\ell,-\mu): \;m  =n- \mu, \; {\rm if}\; (\kappa +1) \mu \in \mathbb Z ,\\
&D(\ell,0): \; m  =n , \; {\rm if}\; \kappa\mu \in \mathbb Z , \;n=0,\pm 1,\pm
2, \dots .
\end{align*}

\section{Discussion and concluding remarks}

We have argued that a consistent pointlike monopole theory with an arbitrary
magnetic charge requires infinite-dimensional representations of the rotation
group, which in general case are multi-valued. Note, that in Dirac theory
`quantization of magnetic charge' follows from the requirement of the wave
function be single-valued. However, the requirement of single-valuedness for a
wave function is not one of the fundamental principles of quantum mechanics,
and having multi-valued wave functions may be allowed until it does not affect
the algebra of observables.

For the single-valued infinite-dimensional representations the generalized
quantization condition is emerged, and depending on the irreducible
representation $D(\ell,\nu)$, has the following form: $(1\pm\kappa)\mu \in
\mathbb Z$ or $\kappa\mu \in \mathbb Z$, where $\kappa$ is the weight of the
Dirac string. In particular cases $\kappa= \pm 1$ and $\kappa =0$ we obtain the
Dirac and Schwinger selectional rules respectively. The obtained quantization
conditions are not mandatory, but once being introduced would give rise to a
new {\em quantum number}, $\nu = \kappa\mu$. This quantum number being related
to a winding number of the map $S^1 \rightarrow U(1)$ has a topological nature,
and can be considered as the {\em topological spin} carried by the Dirac
string.

{\em Concerning experiment.} Recently, in the context of the anomalous Hall
effect, the experimental results providing evidence for the magnetic monopole
in the crystal-momentum space has been reported \cite{FNT}. Besides, in the
literature there are others experimental proposals for the experimental probe
of the `fictitious' monopoles that are appeared in the context of the Berry
phase \cite{Br,ZLS,Hal,FP,SR,MSN,BM}. The experimental observation of the
anomalous scattering on the trapped $\Lambda$ atom with induced magnetic
monopole \cite{ZLS} would provide a direct confirmation of non-quantized Dirac
monopole. The other possible experiment, where the existence of non-quantized
Dirac string in the Berry phase of anisotropic spin systems would be proved,
has been proposed in \cite{Br}.

Other type of the possible experiments, where the Dirac string could be
observed, is related to the scattering on magnetic charge. As was shown in
\cite{Boul,Schetal}, the Dirac string is not observable from the standpoints of
quantum-mechanical scattering processes, if $2\mu =n$. However, the situation
is quite different for the scattering on the monopole with an arbitrary
magnetic charge. In particular, for small scattering angles $\theta$ the
scattering amplitude $f(\theta)$ behaves as $f(\theta) \sim
\sin^{-2}(\theta/2)$, if the Dirac quantization rule holds, and this is agree
with the classical results. In the case of an arbitrary $\mu$ we have
$f(\theta) \sim \sin^{-2\mu}(\theta/2)$, and our numerical results show that
this small angle approximation is good enough up to about $3\pi/4$
\cite{NF2,NF4}. Note, that this novel effect, being produced by a singular
string, is a pure quantum gauge-invariant phenomena, and therefore the
orientation of the Dirac string is not observable in the scattering experiments
(see also \cite{Boul}).

We close with some comments on observability of the Dirac string in the
Aharonov-Bohm (AB) effect \cite{AB}. As is known the AB effect is appeared in
quantum interference between two parts of a beam of charged particles, say
electrons with charge $e$, passing by an infinite long solenoid. In spite of
the fact that the magnetic field $\mathbf B $ outside the solenoid is equal to
zero, it produces an interference effect at the point $Q$ of the screen. A
relative phase shift $ \Delta \varphi$ is given by
\begin{equation}
\Delta \varphi = e\oint_{\mathcal C}\mathbf A \cdot d \mathbf r = e \Phi,
\label{eq16}
\end {equation}
where $\Phi$ is the total magnetic flux through the solenoid. The condition for
the absence of observable AB effect is $e\Phi = 2\pi n, \; n\in \mathbb Z$.

What makes difference between the infinite long solenoid and Dirac's string is
that the latter can be moved by a singular gauge transformation. This means
that the monopole string can not be observed in AB experiment, if singular
gauge transformations are allowed. Thus, the absence of the AB effect for the
Dirac string has a crucial significance for a consistent magnetic monopole
theory.

Let us assume that the beam passes in the upper half of the space divided by
the plane $z=0$. Then the contribution of the string $S^\kappa_{\mathbf n}$ to
the relative phase shift of the wave function at the point $Q$ is found to be
\begin{align}
\Delta \varphi_{+} = 2\pi(1+\kappa)\mu \label{eq11}
\end{align}
and, if the beam passes in the lower half-space ($z < 0$), one has
\begin{align}
\Delta \varphi_{-} = 2\pi(\kappa-1)\mu \label{eq11a}
\end{align}

It follows from the Eqs.(\ref{eq11}) and (\ref{eq11}) the absence of AB effect
when $(1+\kappa)\mu$ and $(1-\kappa)\mu$ are integers. In the case of $\kappa
\neq 0$, this yields immediately the following conditions: $2\mu \in \mathbb
Z$, that is the celebrated Dirac rule, and quantization of the string weight,
namely, $2\kappa\mu \in \mathbb Z$. If $\kappa =0$, one obtains the Schwinger
quantization condition, $\mu \in \mathbb Z$.

At first sight our results are in contradiction with the AB experiment. To
clarify issue let us recall that the phase shift (\ref{eq16}) arises as result
of the parallel translation of wave function along the contour $\mathcal C$
surrounding the Dirac string. It is known that for the generators of
translations the Jacobi identity fails and for the finite translations one has
\cite{Jac,Gr}
\begin{align}
\bigl(U_{\mathbf a}U_{\mathbf b}\bigr)U_{\mathbf c}
=\exp(i\alpha_3(\mathbf r ;\mathbf a, \mathbf b,\mathbf c)) U_{\mathbf
a}\bigl(U_{\mathbf b}U_{\mathbf c}\bigr) \label{as}
\end{align}
where $\alpha_s$ is the so-called {\em three cocycle}, and $\alpha_3= 4\pi
\mu\,\mod 2\pi \mathbb Z$, if the monopole is enclosed by the simplex with
vertices $(\mathbf r,\mathbf r +\mathbf a, \mathbf r+ \mathbf a +\mathbf
b,\mathbf r+ \mathbf a+ \mathbf b +\mathbf c)$ and zero otherwise \cite{Jac}.
For the Dirac quantization condition being satisfied $\alpha_3 = 0 \mod 2\pi
\mathbb Z$, and (\ref{as}) provides an associative representation of the
translations, in spite of the fact that the Jacobi identity continues to fail.
Thus, we see that the AB effect requires more careful analysis, if we assume
existence of an arbitrary monopole charge.

The emerging difficulties in explanation of the AB effect may be removed by
introducing nonassociative path-dependent wave function $\Psi(\mathbf r;
\gamma)$, that provides the absence of the AB effect for an arbitrary magnetic
charge \cite{N1a,N3}.

\section*{Acknowledgements}
 This work was partly supported by SEP-PROMEP (Grant No. 103.5/04/1911).

\appendix

\section{Matrix elements of representations}\label{sec:A}

Here we perform computation of the matrix elements of
representations has been discussed in the text.\\

{\bf Representations unbounded from above and below}. The matrix elements of
the representation $D(\ell, \nu)$ are defined as the coefficients of the
expansion
\begin{align}
T_g|\ell, m'\rangle = \sum {\mathcal D}^{(\ell,\nu)}_{m m'}(g)|\ell,
m\rangle\label{ap1_},
\end{align}
where (see Eq.(\ref{rep_2}))
\begin{align}
&\langle z | T_g|\ell, m'\rangle = {\mathcal N}_{m'}(\bar{\alpha} + \beta
z)^{\ell +m'} ( \alpha z - \bar{\beta})^{\ell-m'}. \label{ap2}
\end{align}
The matrix elements $ {\mathcal D}^{(\ell,\nu)}_{m m'}(g)$ can be obtained as
follows:
\begin{align}
{\mathcal D}^{(\ell,\nu)}_{m m'}(g) =\frac{{\mathcal N}_{m'}}{{\mathcal
N}_m}\frac{1}{2\pi i} \oint_{\gamma}(\bar{\alpha} + \beta z)^{\ell +m'} (
\alpha z - \bar{\beta})^{\ell-m'}z^{m- \ell -1} d z \label{ap3},
\end{align}
where
\begin{align}
\frac{{\mathcal N}_{m'}}{{\mathcal N}_m}= \bigg(\frac{\Gamma(\ell -m
+1)\Gamma(\ell + m + 1)}{\Gamma(\ell -m' +1)\Gamma(\ell + m' + 1)}\bigg)^{1/2}
 \label{ap4}.
\end{align}
Making change of variables as follows: $z= (|\beta|^2 t - 1)/{\alpha\beta t}$,
we find
\begin{align}
{\mathcal D}^{(\ell,\nu)}_{m m'}(g) = - \frac{{\mathcal N}_{m'}}{{\mathcal
N}_m}\cdot\frac{\beta^{m'-m}} {\alpha^{m'+m}}\frac{1} {2\pi i} \oint_{C}(-t)^{-
\ell -m -1}(1-t)^{\ell +m'} ( 1 - t|\beta|^2 )^{m -\ell-1} d t \label{ap5},
\end{align}
where we assume $m'-m$ being an positive integer.

Using the integral representation of the hypergeometric function \cite{BE,LL}
\begin{align}
F(a,b,c;z) = - \frac{\Gamma(1-a)\Gamma(c)}{\Gamma(c-a)}\cdot\frac{1} {2\pi i}
\oint_{C}(-t)^{a -1}(1-t)^{c-a-1} ( 1 - t z )^{-b} d t , \label{ap6}
\end{align}
we obtain
\begin{align}
{\mathcal D}^{(\ell,\nu)}_{m m'}(g)= \beta^{m'-m}\alpha^{-(m+m')}C^{\ell}_{mm'}
F(-\ell-m,\ell -m +1, m'-m+1;|\beta|^2), \qquad m' > m,
 \label{ap7}
\end{align}
where
\begin{align}
C^{\ell}_{mm'}=\frac{1}{(m'-m)!}\bigg(\frac{\Gamma(\ell - m
+1)\Gamma(\ell + m' + 1)}{\Gamma(\ell + m +1)\Gamma(\ell - m' +
1)}\bigg)^{\frac{1}{2}} \label{ap2a}
\end{align}

As known the hypergeometric series
\begin{align}
F(a,b,c;z)=
\frac{\Gamma(c)}{\Gamma(a)\,\Gamma(b)}\sum_{n=0}^{\infty}\frac{\Gamma(a+n)\Gamma(b+n)}{\Gamma(c+n)}
\frac {z^{n}}{n!}
 \label{ap7b}
\end{align}
is not defined when $c=-p$ ($p=0,1,2,\dots$), and for $c$ being
equal to $-p$, one has \cite{abr}
\begin{align}
\lim_{c\rightarrow -p}\frac{1}{\Gamma(c)}F(a,b,c;z)=
\frac{\Gamma(a+p+1)\,\Gamma(b+p+1)}{(p+1)!\,\Gamma(a)\,\Gamma(b)}z^{p+1}F(a+p+1,b+p+1,p+2;z)
 \label{ap7a}
\end{align}
Applying (\ref{ap7a}) to (\ref{ap7}), we obtain in the case $m' < m$
the following expression for the matrix elements:
\begin{align}
{\mathcal D}^{(\ell,\nu)}_{m m'}(g)= \beta^{m'-m}\alpha^{-(m+m')}C^{\ell}_{m'm}
F(-\ell-m',\ell -m' +1, m-m'+1;|\beta|^2), \quad m' < m.
 \label{ap8}
\end{align}

In terms of the usual Cayley-Klein parametrization \cite{Vil,Wig_1,and}
\begin{align}
\alpha = e^{i\varphi/2}\cos\frac{\theta}{2}\;e^{i\psi/2}, \quad \beta
=ie^{i\varphi/2}\sin\frac{\theta}{2}\;e^{-i\psi/2}, \quad
\end{align}
with Euler's angles being $(\varphi,\theta,\psi)$, the formulae (\ref{ap7}) and
(\ref{ap8}) read
\begin{align}
{\mathcal D}^{(\ell,\nu)}_{m m'}(g)= &e^{-im\varphi}e^{-im'\psi}(-i)^{m
-m'}z^{(m'-m)/2} (1-z)^{-(m+m')/2} C^{\ell}_{mm'} \nonumber \\
& \times F(-\ell-m,\ell-m +1, 1+ m'-m;z),  \; {\rm if}\; m' > m \label{ap8b} \\
{\mathcal D}^{(\ell,\nu)}_{m m'}(g)= &e^{-im\varphi}e^{-im'\psi}(-i)^{m'
-m}z^{(m-m')/2} (1-z)^{-(m+m')/2} C^{\ell}_{m' m} \nonumber \\
&\times F(-\ell-m',\ell -m' +1, 1+ m-m';z),  \; {\rm if}\; m' < m \label{ap8c},
\end{align}
where we set $z=\sin^2(\theta/2)$.\\

{\bf Representations bounded above}. The matrix elements of the representation
$D^{-}(\ell,\nu)$ are defined as the coefficients of the expansion
\begin{align}
T_g|\ell, n'\rangle = \sum {\mathcal D}^{-(\ell)}_{nn'}(g)|\ell,
n\rangle\label{ap1a_},
\end{align}
and may be obtained from (\ref{ap8b}) and (\ref{ap8c}) by
substitution $m=\ell - n$. The computation yields
\begin{align}
{\mathcal D}^{-(\ell)}_{n n'}(g)=& e^{i(n- \ell)\varphi} e^{i(n'-
\ell)\psi}(-i)^{n- n'} z^{(n'-n)/2} (1-z)^{(n+n' -2\ell)/2} C^{-\ell}_{nn'}
\nonumber \\
& \times F(n'-2\ell,n' +1,1+ n'-n;z), \; {\rm if}\; n' > n \label{ap10_}\\
{\mathcal D}^{-(\ell)}_{n n'}(g)=& e^{i(n-\ell)\varphi} e^{i(n'-
\ell)\psi}(-i)^{n'- n} z^{(n-n')/2} (1-z)^{(n+n' -2\ell)/2} C^{-\ell}_{n'n}
\nonumber \\
& \times F(n-2\ell,n +1,1+ n-n';z), \; {\rm if}\; n' < n \label{ap11_}
\end{align}
where
\begin{align}
C^{-\ell}_{nn'}=\frac{1}{(n'-n)!}\bigg(\frac{\Gamma(n' +1)\Gamma(2\ell -n +
1)}{\Gamma(2\ell -n' +1)\Gamma(n + 1)}\bigg)^{\frac{1}{2}}
\end{align}

Note that the matrix element ${\mathcal D}^{+(\ell)}_{n n'}(g)$ given in
Eq.(\ref{ap10_}) may be rescat in terms of Jacobi's polynomials
$P_n^{(\alpha,\beta)}(x)$ by the application of the following relations
\cite{abr}
\begin{align}
&F(a,b,c;z)= (1-z)^{c-a-b}F(c-a,c-b,c;z)\label{ap11a} \\
&F(-n,\alpha + \beta +n +1, \alpha +1;z)=
\frac{\Gamma(n+1)\,\Gamma(\alpha + 1)}{\Gamma(n+\alpha + 1)}\,
P_n^{(\alpha,\beta)}( 1-2z)\label{ap11b}
\end{align}
to yield
\begin{align}
{\mathcal D}^{-(\ell)}_{n n'}(g)=  &\bigg(\frac{\Gamma(n +1)\Gamma(2\ell -n +
1)}{\Gamma(n' +1)\Gamma(2\ell -n' + 1)}\bigg)^{1/2}e^{i(n- \ell)\varphi}
e^{i(n'- \ell)\psi}(-i)^{n- n'}\nonumber \\
& \times z^{\alpha/2} (1-z)^{\beta/2}P_n^{(\alpha,\beta)}( 1-2z), \; \; {\rm
if}\; n' > n \label{ap10a} \\
{\mathcal D}^{-(\ell)}_{n n'}(g)=  &\bigg(\frac{\Gamma(n' +1)\Gamma(2\ell -n' +
1)}{\Gamma(n +1)\Gamma(2\ell -n + 1)}\bigg)^{1/2}e^{i(n- \ell)\varphi} e^{i(n'-
\ell)\psi}(-i)^{n'- n} \nonumber \\
& \times z^{-\alpha/2} (1-z)^{\beta/2}P_n^{(-\alpha,\beta)}( 1-2z), \; \; {\rm
if}\; n'< n \label{ap10b}
\end{align}
where $\alpha =n'-n, \quad \beta= 2\ell -n-n'$.\\

{\bf Representations bounded below}. The matrix elements of the representation
$D^{+}(\ell,\nu)$ may be obtained from (\ref{ap8b}) and (\ref{ap8c}) by
substitution $m=n -\ell$ and are written as follows:
\begin{align}
{\mathcal D}^{+(\ell)}_{n n'}(g)=& e^{i(\ell-n)\varphi} e^{i(\ell -
n')\psi}(-i)^{n- n'} z^{(n-n')/2} (1-z)^{( 2\ell-n-n')/2} C^{+\ell}_{nn'}
\nonumber \\
& \times F(-n,2\ell -n +1,1+ n'-n;z), \; {\rm if}\; n' > n \label{ap12_}\\
{\mathcal D}^{+(\ell)}_{n n'}(g)= &e^{i(\ell-n)\varphi} e^{i(\ell -
n')\psi}(-i)^{n'- n} z^{(n'-n)/2} (1-z)^{( 2\ell-n-n')/2}
C^{+\ell}_{nn'}\nonumber \\
& \times F(-n',2\ell -n' +1,1+ n-n';z), \; {\rm if}\; n' < n \label{ap13_}
\end{align}
where
\begin{align}
C^{+\ell}_{nn'}=\frac{1}{(n'-n)!}\bigg(\frac{\Gamma(2\ell - n +1)\Gamma(n' +
1)}{\Gamma(n + 1)\Gamma(2\ell - n' +1)}\bigg)^{\frac{1}{2}} \label{ap14}
\end{align}
The matrix elements ${\mathcal D}^{-(\ell)}_{n n'}(g)$ can be written also as
\begin{align}
{\mathcal D}^{+(\ell)}_{n n'}(g)= & \bigg(\frac{\Gamma(n +1)\Gamma(2\ell -n +
1)}{\Gamma(n' +1)\Gamma(2\ell -n' + 1)}\bigg)^{1/2}e^{i(\ell-n)\varphi}
e^{i(\ell -n')\psi}(-i)^{n- n'}\nonumber \\
& \times z^{-\alpha/2} (1-z)^{\beta/2}P_n^{(-\alpha,\beta)}( 1-2z), \; \;
{\rm for}\; n' > n \label{ap13a}\\
{\mathcal D}^{+(\ell)}_{n n'}(g)= &\bigg(\frac{\Gamma(n' +1)\Gamma(2\ell -n' +
1)}{\Gamma(n +1)\Gamma(2\ell -n+ 1)}\bigg)^{1/2} e^{i(\ell-n)\varphi} e^{i(\ell
-n')\psi}(-i)^{n'- n}\nonumber \\
& \times z^{\alpha/2} (1-z)^{\beta/2}P_n^{(\alpha,\beta)}( 1-2z), \; \; {\rm
for}\; n'< n. \label{ap13b}
\end{align}

\section{Some properties of the generalized monopole harmonics}\label{sec:B}

In this section some properties of generalized monopole harmonics,
$\stackrel{{\quad}\pm\;(\mu,n)}{Y_{\ell \pm \mu}}(\theta,\varphi)$, will be
derived. Recall that the functions $\stackrel{{\quad}+\;(\mu,n)}{Y_{\ell -
\mu}}(\theta,\varphi)$ and $\stackrel{{\quad}-\;(\mu,n)}{Y_{\ell +
\mu}}(\theta,\varphi)$ form the basis of the infinite-dimensional
representations $D^{+}(\ell,-\mu)$ and $D^-(\ell,-\mu)$, respectively.  In
terms of the functions
\begin{equation}\label{C1}
Y_n^{(\delta,\gamma)}(u) =C_n\,(1-u)^{\delta/2}(1+u)^{\gamma/2}
P_n^{(\delta,\gamma)}(u),
\end{equation}
where $P_n^{(\delta,\gamma)}(u)$ are the Jacobi polynomials, $u=\cos\theta$ and
the normalization constant being
\begin{align}
C_n=\Bigg(\frac{2 \pi \, 2^{\delta +\gamma +1}\Gamma(n +\delta +1) \Gamma(n
+\gamma +1)}{(2n +\delta +\gamma +1)\Gamma(n+1)\Gamma(n +\delta
+\gamma+1)}\Bigg)^{-1/2} \label{C2}
\end{align}
one has
\begin{align}
&\stackrel{{\quad}-\;(\mu,n)}{Y_{\ell+\mu}}
=e^{i\alpha\varphi}Y_n^{(\alpha,\beta)}(u), \;\;m= \ell -n,\;\ell+\mu \in {\mathbb Z}_{+}\\
&\stackrel{{\quad}+\;(\mu,n)}{Y_{\ell- \mu}}
=e^{i\alpha\varphi}Y_n^{(-\alpha,-\beta)}(u),\; \;m= n-\ell,\;\ell-\mu \in
{\mathbb Z}_{+} \label{C3}
\end {align}
where we set $\alpha = m+\mu$ ($ \alpha \in {\mathbb Z}$) and $\beta = m-\mu$.

It is easy to show that the generalized monopole harmonics satisfy the
orthonormality conditions
\begin{align}
&{\smallsetminus{\hspace{-0.36cm}}\int}\,\overline{\stackrel{{\quad}\mp\;(\mu,n)}{Y_{\ell\pm\mu}}}(\theta,\varphi)
\stackrel{{\quad}\mp\;(\mu,n')}{Y_{\ell'\pm\mu}}(\theta,\varphi)\,d\Omega =
\delta_{\ell\ell'}\eta_{nn'} \\
&{\smallsetminus{\hspace{-0.36cm}}\int}\,\overline{\stackrel{{\quad}\pm\;(\mu,n)}{Y_{\ell\mp\mu}}}(\theta,\varphi)
\stackrel{{\quad}\mp\;(\mu,n')}{Y_{\ell'\pm\mu}}(\theta,\varphi)\,d\Omega = 0
\end{align}
where $\eta_{nn'}= {\rm sgn}(\Gamma(2\ell-n+1)\,\delta_{nn'}$. Using the well
known relation $ \Gamma (z)\Gamma (1-z)= \pi/\sin(\pi z)$ \cite{Leb}, one can
find
\begin{align}
&\eta_{np}= \left\{\begin{array}{l}
\delta_{np},\; {\rm if}\; 2\ell -n >0\\
\delta_{np}(-1)^{n+1}\,{\rm sgn}(\sin2\pi\ell),\; {\rm if}\; n -2\ell >0.
  \end{array}\right.
\end{align}

The orthonormality relations are completed by the completeness conditions
\begin{align}
\sum_{\ell,n,p}\,\overline{\stackrel{{\quad}\mp\;(\mu,n)}{Y_{\ell\pm\mu}}}(\theta,\varphi)\;
\eta_{np}\hspace{-0.25cm}\stackrel{{\quad}\mp\;(\mu,p)}{Y_{\ell\pm\mu}}(\theta',\varphi')
= \frac{1}{\sin^2\theta}\, \delta(\theta-\theta') \delta(\varphi -\varphi')
\end{align}

\subsection*{Addition theorem}

The addition theorem for generalized monopole harmonics can be obtained from
the composition of successive transformations,
\begin{align}
T(g'^{-1})T(g)= T(\tilde g)
\end{align}
where $\tilde g = g'^{-1}g$. Taking into account that $T(g'^{-1})=
T^\dag_\eta (g') $, where $T^\dag_\eta = \eta^{-1}T^\dag \eta$, we
find $T(g')^\dag_\eta\, T(g)= T(\tilde g) $. This yields the desired
addition theorem as follows:
\begin{align}
\sum_{n''} ({\mathcal D}^\dag_{\eta}){}_{nn''}(g') {\mathcal D}_{n''n'}(g) =
\sum_{p,q,n''} \eta_{n''p} \,\overline {\mathcal
D}_{pq}(g')\,\eta^{-1}_{qn}\,{\mathcal D}_{n''n'}(g) ={\mathcal D}_{nn'}(\tilde
g) \label{C9}
\end{align}
In terms of Euler's angles, assuming $g=g(\varphi,\theta,\psi)$,
$g'=g(\varphi',\theta',\psi')$ and $\tilde
g=g(\tilde\varphi,\tilde\theta,\tilde\psi)$, we obtain the well known results
\cite{Schetal,Ed}
\begin{align}
&\cos\tilde\theta = \cos\theta\cos\theta' + \sin\theta\sin\theta'\cos{(\varphi
- \varphi')}\\
&\tilde \psi = \psi +\bar\psi, \quad \tilde\varphi=\bar\psi' -\psi' \label{C10}
\end{align}
where $\bar\psi$ and $\bar\psi' $ being functions of $\theta,\theta'$ and
$\varphi - \varphi'$ are determined from
\begin{align}
&\tan\frac{\bar\psi+\bar\psi'}{2}=\frac{\cos\frac{\theta+\theta'}{2}}
{\cos\frac{\theta-\theta'}{2}}\,\tan\frac{\varphi-\varphi'}{2}\\
&\tan\frac{\bar\psi-\bar\psi'}{2}=-\frac{\sin\frac{\theta+\theta'}{2}}
{\sin\frac{\theta-\theta'}{2}}\,\tan\frac{\varphi-\varphi'}{2} \label{C11}
\end{align}
Note that (\ref{C9}) can be rewritten as follows:
\begin{align}
\overline {\mathcal D}_{nn'}(\tilde g) = \sum_{p,q,n''} \eta_{n''p} \,
{\mathcal D}_{pq}(g')\,\eta^{-1}_{qn}\,\overline {\mathcal D}_{n''n'}(g)=
\sum_{n''}\overline {\mathcal D}_{n''n'}(g) \,{({\mathcal
D}_\eta)}{}_{n''n}(g')\label{C12}
\end{align}

It easy to see that for the generalized monopole harmonics the following
relation holds
\begin{align}
{\stackrel{{\;\;\,}\mp\;(\mu,n)}{Y_{\ell\pm\mu}}}( \theta, \varphi) =
\bigg(\frac{4\pi}{2\ell +1}\bigg)^{-1/2} {\mbox{e}}^{-i\mu\psi}\,\overline
{{\mathcal D}^{\mp(\ell)}_{n \,\ell\pm\mu}}(g) \label{C13}
\end{align}
Now applying (\ref{C13}), we obtain the following addition theorem for the
generalized monopole harmonics:
\begin{align}
{\stackrel{{\;\;\,}\mp\;(\mu,n)}{Y_{\ell\pm\mu}}}(\tilde \theta,\tilde \varphi)
= {\mbox{e}}^{i\mu(\tilde\psi -\psi )}
\sum_{n'}\,\overline{\stackrel{{\quad}\mp\;(\mu,n')}{Y_{\ell\pm\mu}}}(\theta,\varphi)\;
{({\mathcal D}_\eta)}{}_{n'n}(g') \label{C14}
\end{align}
Since $\psi$ is not physical parameter, one can set in $\psi=\psi'=0$, and
inserting $n=\ell\pm\mu$ in (\ref{C14}), we obtain
\begin{align}
{\stackrel{{\quad\;\;\,}\mp\;(\mu,\ell\pm\mu)}{Y_{\ell\pm\mu}}}(\tilde
\theta,\tilde \varphi) = \bigg(\frac{4\pi}{2\ell +1}\bigg)^{-1/2}
{\mbox{e}}^{i\mu\tilde\psi
}\sum_{n',n''}\,\overline{\stackrel{{\quad}\mp\;(\mu,n')}{Y_{\ell\pm\mu}}}(\theta,\varphi)
\;\eta_{n'n''}{\hspace{-0.3cm}}{\stackrel{{\quad}\mp\;(\mu,n'')}{Y_{\ell\pm\mu}}}(\theta',\varphi')\label{C15}
\end{align}

Note that consideration of the representation $D^{\pm}(\ell,\mu)$ and
associated generalized harmonics $\stackrel{{\quad}\pm(-\mu,n)}{Y_{\ell \pm
\mu}}(\theta,\varphi)$ can be done in a similar way. In fact, the corresponding
relations can be obtained from (\ref{C3}) - (\ref{C15}) by substitution $\mu
\rightarrow -\mu$.

\section{General solution of the hypergeometric equation and magnetic monopoles}\label{sec:C}

Here we consider the general situation with a string of an arbitrary weight
$\kappa$. For simplicity, we assume the string to be a straight line along the
axes $z$, then
\begin{align}
\mathbf{A}^\kappa_{\mathbf n} = \frac{1+ \kappa}{2}{\mathbf A}_{\mathbf n} +
\frac{1- \kappa}{2}{\mathbf A}_{-\mathbf n}, \label{eqB1b}
\end{align}
and the generators of the rotation group are written as follows:
\begin{align}
J_{\pm}= &\;e^{\pm i\varphi}\bigg(\pm\frac{\partial}{\partial\theta}
+i\cot\theta \frac{\partial}{\partial\varphi} - \frac{(\mu + \gamma\cos
\theta)\sin\theta}{1-\cos^2\theta} \bigg) , \quad
J_3=-i\frac{\partial}{\partial\varphi}+ \gamma \nonumber \\
\mathbf{J}^2 = & -\frac{1}{\sin\theta}\frac{\partial}{\partial\theta}
\left(\sin\theta\frac{\partial}{\partial\theta}\right) -
\frac{1}{\sin^2\theta}\frac{\partial^2}{\partial\varphi^2} + \frac{(\gamma +
\mu\cos\theta)^2}{\sin^2\theta} - \frac{2i(\gamma + \mu\cos\theta)}
{\sin^2\theta}\frac{\partial}{\partial\varphi} + \mu^2 \label{eqB1a}
\end{align}
where $\gamma = \kappa \mu$.

The general solution of the Schr\"odinger equation (\ref{eq7a}) has the form:
\begin{align}
&Y\propto {e}^{-i\gamma\varphi} {e}^{im\varphi}
z^{(m+\mu)/2}(1-z)^{(m-\mu)/2}\big (\alpha_1 Y_1(z)+\alpha_2 Y_2(z)\big)
\label{eqsB1a}
\end{align}
where $\alpha_1$ and $\alpha_2$ are arbitrary constants, and without loss of
generality one can set $\alpha_1 +\alpha_2 =1$. The functions $Y_1(z)$ and
$Y_2(z)$ are linearly independent solutions of the hypergeometric equation
\begin{align}
z(1 - z)\frac{d^2 w}{dz^2} + \left(c - (a + b + 1)z\right)\frac{d
w}{dz} - abz = 0 \label{B0}
\end{align}
with $a = m - \ell, \; b = m + \ell + 1$ and $ c = 1 + m + \mu$.

In the neighborhood of the singular points $z=0,1$ two linearly
independent solutions of the hypergeometric equation are written
respectively as \cite{abr}
\begin{align}
w_{1(0)}= &F(a, b, c; z) = (1 - z)^{c - a - b}F(c - a, c - b, c ; z)
\label{B1} \\
w_{2(0)}= &z^{1 - c}F(a-c+1,b-c+1, 2- c; z)=z^{1 - c}(1 - z)^{c - a - b}F(1 - a, 1 - b, 2 - c;z) \label{B2} \\
w_{1(1)}= &F(a, b,a+b+1- c;1- z) =  z^{1-c}F(1+b-c, 1+a-c,a+b+1-c ;1- z) \label{B3}\\
w_{2(1)} =&(1 - z)^{c - a - b}F(c-b,c-a, c-a-b +1;1- z)\nonumber \\
 = &z^{1 - c}(1 - z)^{c - a - b} F(1-a,1-b,c-a-b+1; 1-z)\label{B4}
\end{align}

If one of the numbers $a,b,c-a, c-b$ is integer, then one of the hypergeometric
series (\ref{B1}) -- (\ref{B4}) terminates and the respective solution has the
form $w= z^\alpha(1-z)^\beta P_n(z)$, where $ P_n(z)$ is a polynomial of degree
$n$.

In the text (Sec. IV) we consider the linear combination $\alpha_1 w_{1(0)} +
\alpha_2 w_{1(1)}$. Using the transformation formula
\begin{align}
F(a, b, c; z) = &\frac{\Gamma(c)\Gamma(c - a - b)}{\Gamma(c - a)\Gamma(c -
b)}F(a, b, a + b - c + 1; 1 - z) \nonumber \\
+ &\frac{\Gamma(c)\Gamma(a + b - c)}{\Gamma(a)\Gamma(b)}(1 - z)^{c - a - b}F(c
- a, c - b, c - a - b + 1; 1 - z),
\end{align}
for instance, one can show that $\alpha_1 w_{1(0)} + \alpha_2 w_{1(1)}$ may be
transformed to $\tilde\alpha_1 w_{1(0)} + \tilde\alpha_2 w_{2(0)}$. This
support the claim that any pair of the solutions (\ref{B1}) - (\ref{B4}) can by
used to obtain the general solution $Y$ of the hypergeometric equation
(\ref{B0}). In explicit form there is
\begin{align}
Y = \alpha_1 F(a, b, c; z) + \alpha_2 F(a, b, a + b - c + 1; 1 - z)
\label{B5}
\end{align}
and in case of Dirac monopole problem one has
\begin{equation}
Y=\alpha_1  F(m-\ell, m+\ell+1,1+ m -\mu; 1-z) + \alpha_2F(m-\ell,
m+\ell+1,1+ m +\mu; z) \label{B6}
\end{equation}

From the requirement of the invariance of the obtained solution with respect to
the generalized parity transformation $ \mu \rightarrow -\mu, \mathbf r
\rightarrow -\mathbf r $ (see Sec. III ), it follows $\alpha_1(\mu)=
\alpha_2(-\mu) $. Now, taking into account the condition $\alpha_1 +\alpha_2
=1$, Eq. (\ref{eqB1b}) and behavior of the hypergeometric function at the
points $z=0$ and $z=1$, one can find that $\alpha_1=(1+\kappa)/2$ and
$\alpha_2=(1-\kappa)/2$. This yields
\begin{equation}
Y= \frac{1+\kappa}{2}F(m-\ell, m+\ell+1,1+ m -\mu; 1-z)
+\frac{1-\kappa}{2}F(m-\ell, m+\ell+1,1+ m +\mu; z) \label{B7}
\end{equation}
This implies also, that $\alpha_1 - \alpha_2$ may be identified as the weight
of the Dirac string, namely, $\kappa = \alpha_1 - \alpha_2 $.

Finally we find that the general solution (\ref{eqB1a}) may be written as
follows:
\begin{align}
Y\propto \;&{e}^{i(\alpha_2 - \alpha_1)\mu\varphi} {e}^{im\varphi}
z^{(m+\mu)/2}(1-z)^{(m-\mu)/2}\big (\alpha_1  F(m-\ell, m+\ell+1,1+ m -\mu;
1-z) \nonumber \\
&+ \alpha_2F(m-\ell, m+\ell+1,1+ m +\mu; z)\big) \nonumber
\end{align}
where $\alpha_1 + \alpha_2 =1$.

Let us consider the case $\alpha_1=1$ and $\alpha_2=0$ being associated with
the Dirac string passing from the origin of coordinates to $\infty$. The
corresponding solution of the Schr\"odinger equation is given by
\begin{align}
 Y_{1,\ell}^{(\mu,m)}\propto {e}^{-i\mu\varphi}
{e}^{im\varphi} z^{(m+\mu)/2}(1-z)^{(m-\mu)/2}F(m-\ell, m+\ell+1,1+ m -\mu;
1-z)
\end{align}
should be singular at the north pole ($z=0$), where the string crosses the
sphere, and regular at the south pole ($z=1$).

The computation yields $c-a-b = -(m+\mu)$, and according to the general
properties of the hypergeometric function one can see that
$Y_{1,\ell}^{(\mu,m)}$ goes to infinity like $z^{-|m + \mu|/2}$, $z \rightarrow
0$. Now taking into account the well known relation
\begin{align}
\lim_{c\rightarrow -p}\frac{1}{\Gamma(c)}F(a,b,c;z)=
\frac{\Gamma(a+p+1)\,\Gamma(b+p+1)}{(p+1)!\,\Gamma(a)\,\Gamma(b)}z^{p+1}F(a+p+1,b+p+1,p+2;z)
 \label{B8}
\end{align}
we find that the obtained solution has no singularity at $z=1$, if $m - \mu$ is
an integer.

In a similar way, choosing $\alpha_1=0$ and $\alpha_2=1$, we find that the
corresponding solution
\begin{align}
 Y_{-1,\ell}^{(\mu,m)}\propto {e}^{i\mu\varphi}
{e}^{im\varphi} z^{(m+\mu)/2}(1-z)^{(m-\mu)/2}F(m-\ell, m+\ell+1,1+ m +\mu; z)
\end{align}
being singular at the south pole($z=1$) is regular at $z=0$, if $m + \mu$ is an
integer. Note that in both cases the solutions are given by single-valued
functions.

Finally, the choice of $\alpha_1=\alpha_2=0$ being related to the Schwinger
string, yields
\begin{align}
Y_{0,\ell}^{(\mu,m)}\propto \;&{e}^{im\varphi}
z^{(m+\mu)/2}(1-z)^{(m-\mu)/2}\big
(F(m-\ell, m+\ell+1,1+ m -\mu; 1-z)  \nonumber \\
&+ F(m-\ell, m+\ell+1,1+ m +\mu; z)\big).
\end{align}
By requiring $Y_{0,\ell}^{(\mu,m)} $ be a single-valued function, we obtain $m
\in \mathbb Z$.

\providecommand{\href}[2]{#2}\begingroup\raggedright\endgroup

\end{document}